# A companion to "Non-Kähler Deformed Conifold, Ultra-Violet Completion and Supersymmetric Constraints in the Baryonic Branch"

Jake Elituv

*Ernest Rutherford Physics Department, McGill University*
*3600 rue University, Montréal, Québec, Canada H3A 2T8*

E-mail: `jake.elituv@mail.mcgill.ca`

ABSTRACT: In the appropriate limit, a type IIB string theory setup involving D3 branes, wrapped D5 branes, and fluxes on a conifold generally leads to a supergravity background involving a warped version of the conifold with fluxes. We study the supergravity dual of the baryonic branch of the Klebanov Strassler theory by writing down a very general conifold metric—the non-Kähler resolved warped-deformed conifold—and a general set of fluxes that satisfy the supergravity equations of motion, and derive the necessary constraints that allow the geometry to be dual to an $\mathcal{N}=1$ supersymmetric gauge theory in $3+1$ dimensions. These backgrounds encompass known solutions, such as the KS, MN and Butti et al. models, but the added layer of generality can lead to a larger class of gauge-gravity dualities. We also present many consistency checks that validate our background matches known cases for certain values in our parameter space. This is a companion paper to arXiv:1805.03676 [hep-th] covering the section 'IR physics, dualities, and supersymmetry'.

# Contents

1# Chapter 1

# Introduction

The AdS/CFT correspondence [7] is a powerful new tool at physicists' disposal that relates two seemingly unrelated theories: supergravity and supersymmetric gauge theory. One of the main reasons the correspondence is so powerful is that the strongly coupled regime of one theory is dual to the weakly coupled regime of the other; so the correspondence allows us to use perturbative techniques to study non-perturbative properties.

AdS/CFT in its pure form relates superconformal field theories (SCFT) to supergravity theories living on the direct product of an anti-de Sitter (AdS) space and some compact manifold. From the gauge theory point of view, there are physical systems that are modelled well with CFTs such as many types of scale invariant condensed matter systems [15]. However, there are two main features of SCFTs that do not model many physical systems well: the large amount of supersymmetry (SUSY) that needs to be imposed and conformal symmetry. Experimentally, SUSY has yet to be observed and is highly constrained [8], so it is appealing to study theories beyond the Standard Model that require minimal SUSY. Furthermore, any physical system with massive degrees of freedom or RG flow do not have scale invariance away from RG fixed points; to study such non-conformal theories in the strong coupling regime using gauge/gravity duality requires some sort of non-AdS/non-CFT correspondence.

One way examples of non-AdS/non-CFT appear is when we consider type IIB string theory setups on conifold geometries. The six-dimensional conifold, loosely speaking, is a cone-like surface with metric $ds_6^2 = dr^2 + r^2 d\Omega^2$ that is topologically $\mathbb{R}^+ \times \Omega_5$, where $\Omega_5$ is some compact 'base manifold'. The base manifold describes the cross section of the conifold geometry at fixed radius $r$. The singular conifold, which will be defined more precisely in 2.2, has a fixed base manifold known as $T^{1,1}$ which is topologically $S^2 \times S^3$. As $r \to 0$, the 2-sphere and 3-sphere of the singular conifold both shrink to a point and there is a conical singularity, hence its name. The deformed conifold has its 3-sphere remain at finite size at the tip, and the resolved conifold has its 2-sphere remain at finite size at the tip; both will be described in more detail in 2.2.

One of the most popular models of this type, the Klebanov-Strassler model (KS) [6], involves wrapping D5 branes around the vanishing 2-sphere of the singular conifold at the tip along with placing D3 branes at the tip. In the appropriate limit, the resulting gauge theory has $\mathcal{N} = 1$ supersymmetry and RG flow with confinement, making it very similar to QCD. In the supergravity limit of the KS model, the singular conifold geometry becomes a Kähler warped-deformed conifold with fluxes [6][1]. A slightly modified version of the KS model been used, for example, in [9] to study properties of strongly coupled QCD. Other $\mathcal{N} = 1$ models include the Maldacena-Núñez model (MN) [10] and the Vafa model [11] whose supergravity descriptions are a non-Kähler deformed conifold with fluxes and a non-Kähler resolved conifold with fluxes, respectively.

A pattern among the models [6][11][10] is that they are all physically motivated with brane constructions in type IIB string theory and the geometries described in their supergravity regimes are elements of the set of non-Kähler warped-deformed-resolved conifolds with fluxes (in the sense that Kähler metrics are a subset of non-Kähler metrics). So one may wonder, working backwards from the supergravity picture, whether there is a set of conditions on a generic non-Kähler warped-deformed-resolved conifold with fluxes such that the theory is dual to an $\mathcal{N} = 1$ supersymmetric gauge theory. Dasgupta et. al. derived the set of conditions for the case of a generic non-Kähler warped-resolved conifold with fluxes in [4]. Here we will derive the full set of supersymmetry constraints in the more general case that allows both non-Kähler resolution and deformation. These results are based on the section 'IR physics, dualities, and supersymmetry' in [1].

---

[1]The concepts of Kähler metrics and warped conifolds will be discussed in 2.1

# Chapter 2

# Background material

## 2.1 Complex differential geometry

We will review the relevant information about complex differential geometry that we will need for our later calculations. This overview closely follows the relevant chapter in [12].

A complex manifold of complex dimension $n$ is defined similarly to a real manifold except that it uses complex coordinates $(z_1, \ldots, z_n, \overline{z}_1, \ldots, \overline{z}_n)$ instead of real ones and the transition functions between charts are required to be biholomorphic[1] instead of just differentiable.

We can define differential forms on complex manifolds as follows:

$$A_{p,q} = \frac{1}{p!q!} A_{a_1 \cdots a_p \bar{b}_1 \cdots \bar{b}_q} dz^{a_1} \wedge \cdots \wedge dz^{a_p} \wedge d\bar{z}^{\bar{b}_1} \wedge \cdots \wedge d\bar{z}^{\bar{b}_q}. \tag{2.1}$$

A generic complex differential form is labelled by two integers $(p, q)$ if it has $p$ holomorphic basis one forms $dz^i$ and $q$ anti-holomorphic basis one forms $d\bar{z}^i$, as shown above.

---
[1] A function $f$ is biholomorphic if both $f$ and $f^{-1}$ are holomorphic

Every complex manifold has the property that it admits a globally defined tensor $\mathbf{J}$, called the complex structure, satisfying

$$\mathbf{J}_a{}^b = i\delta_a{}^b, \quad \mathbf{J}_{\bar{a}}{}^{\bar{b}} = -i\delta_{\bar{a}}{}^{\bar{b}}, \quad \mathbf{J}_a{}^{\bar{b}} = \mathbf{J}_{\bar{a}}{}^b = 0. \tag{2.2}$$

We can then ask the following question: when is a real manifold a complex manifold? All complex manifolds are necessarily real manifolds; this can be seen by decomposing

$$z_j = x_j + iy_j, \quad \bar{z}_j = x_j - iy_j \tag{2.3}$$

and letting $(x_1, \ldots, x_n, y_1, \ldots, y_n)$ be the coordinates on the real manifold. From this we can easily see that a real manifold necessarily needs to be of even dimension $2n$ if is to also be complex.

The first requirement a real manifold needs to have to be a complex manifold is a tensor, known as an almost complex structure, satisfying

$$\mathbf{J}_m{}^n \mathbf{J}_n{}^p = -\delta_m{}^p. \tag{2.4}$$

Furthermore, for the real manifold to be complex, this tensor must be a complex structure, defined in (2.2). It can be shown that an almost complex structure is a complex structure when the Nijenhuis tensor, defined as

$$N^p{}_{mn} = \mathbf{J}_m{}^q \partial_{[q} \mathbf{J}_{n]}{}^p - \mathbf{J}_n{}^q \partial_{[q} \mathbf{J}_{m]}{}^p, \tag{2.5}$$

vanishes [12].

An almost complex structure describes the way we combine $2n$ real coordinates into $n$ complex ones. For example, we assume a specific almost complex structure when we take two real coordinates $x$ and $y$ and combined them into the complex coordinate $z = x + iy$; in our notation we will call this an almost complex structure $(i)$. We could just as easily use the combination $z = x + \sigma i y$, for some real number $\sigma$; we call this an almost complex structure $(i\sigma)$. We will be considering six dimensional real manifolds, so the almost complex structures we will use will be in the form $(i, i, i)$, for example, which corresponds to the case of (2.3), with $j = 1, 2, 3$.

The conifold geometries we will be studying have SU(3) structure [13]. Manifolds with SU(3) structure have a globally defined $(3, 0)$ form, which we will call the holomorphic $(3, 0)$ form $\Omega$ [13]; as we will see in 3, we can construct $\Omega$ by wedging together the three holomorphic complex vielbeins. The vanishing of the Nijenhuis tensor (2.5) is equivalent to the holomorphic $(3, 0)$ form being closed [13]:

$$d\Omega = 0. \tag{2.6}$$

So to determine whether a conifold geometry is complex one needs to show its holomorphic $(3, 0)$ form is closed.

Another quantity we can define on complex manifolds is the Kähler form or fundamental $(1, 1)$ form $J$ defined as

$$J = ig_{a\bar{b}} dz^a \wedge d\bar{z}^{\bar{b}}. \tag{2.7}$$

When this form is closed ($dJ = 0$) the manifold is called Kähler.

Summarizing the relevant section of [13], a supergravity setup is dual to an $\mathcal{N} = 1$ theory only if the geometry is a complex manifold and its complex 3-form flux $\mathbf{G}_3 = \mathbf{F}_3 - ie^{-\phi}\mathbf{H}_3$ is a primitive[2] $(2, 1)$ form. So to determine the supersymmetry constraints on a conifold geometry we must check the following:

---

[2]$\mathbf{G}_3$ is primitive if $J \wedge \mathbf{G}_3 = 0$ [5]

1. Compute the holomorphic $(3,0)$ form $\Omega$ and compute the conditions for $\Omega$ to be closed

2. Compute the complex 3-form flux $\mathbf{G}_3$ and expand it in a basis of holomorphic and anti-holomorphic basis one-forms as in (2.1) and compute the conditions that set all the non-$(2,1)$ pieces to 0

3. Verify that the $(2,1)$-form solution of $\mathbf{G}_3$ satisfies $J \wedge \mathbf{G}_3 = 0$.

## 2.2 Geometry of the Conifold

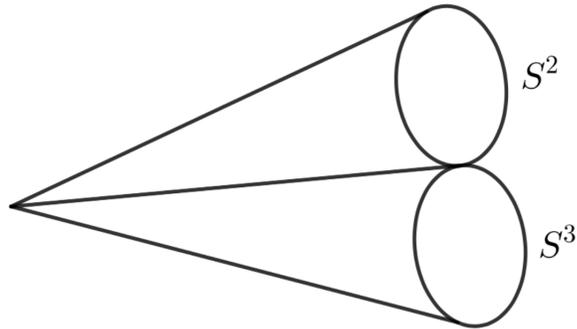

FIGURE 2.1: The singular conifold

This section closely follows the references [2] and [5]. We define the singular conifold as a subspace of $\mathbb{C}^4$ defined by

$$f(w) = \sum_{i=1}^{4} w_i^2 = 0, \tag{2.8}$$

where $w = (w_1, \ldots, w_4) \in \mathbb{C}^4$. The space is 'cone-like' because it is spanned by a set of lines (since if $w$ is a solution, $\lambda w$ is also a solution) that all converge to one point ($w = 0$). The space is singular because at $w = 0$, $df|_{w=0} = 2w_i dw^i|_{w=0} = 0$.

We would like to construct a Kähler metric on the singular conifold. We consider the ansatz

$$ds^2 = dr^2 + r^2 d\Omega^2, \tag{2.9}$$

where $d\Omega$ is the angular part of the metric we would like to determine. We begin by isolating the angular portion of the singular conifold by considering the set of $w$ that both satisfy (2.8) and

$$|w|^2 = \sum_{i=1}^{4} |w_i|^2 = r^2. \tag{2.10}$$

Expanding $w = x + iy$, (2.8) and (2.10) imply

$$|x|^2 = \left(\frac{r}{\sqrt{2}}\right)^2, \quad |y|^2 = \left(\frac{r}{\sqrt{2}}\right)^2, \quad x \cdot y = 0, \tag{2.11}$$

which topologically describes an $S^2$ fibered over an $S^3$. Arguments in [2] show that the unique Ricci flat metric compatible with both the topology $S^2 \times S^3$ and the complex structure implied by our expansion of $w$ is the metric $T^{1,1}$ defined by

$$d\Omega^2_{T^{1,1}} = \frac{1}{9}(d\psi + \cos\theta_1 d\phi_1 + \cos\theta_2 d\phi_2)^2 + \frac{1}{6}\sum_{i=1}^{2}(d\theta_i^2 + \sin^2\theta_i d\phi_i^2). \tag{2.12}$$

To obtain the deformed conifold we modify the defining equation (2.8) to the following:

$$f(w) = \sum_{i=1}^{4} w_i^2 = \mu^2, \tag{2.13}$$

where $\mu^2$ is known as the deformation parameter. To obtain the resolved conifold, we first

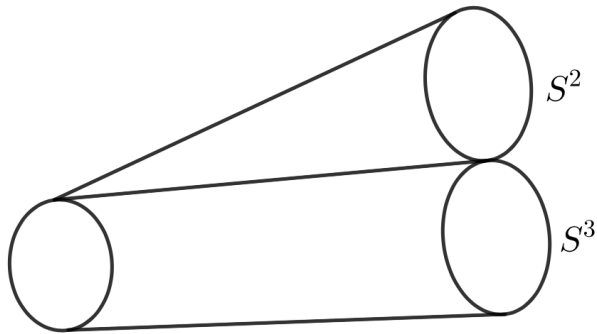

FIGURE 2.2: The deformed conifold

make the change of variables

$$x = w_1 + iw_2, \quad y = w_2 + iw_1$$
$$u = w_3 - iw_4, \quad v = w_4 - iw_3 \tag{2.14}$$

so that (2.8) becomes

$$0 = \sum_{i=1}^{4} w_i^2 = \left(\frac{x-iy}{2}\right)^2 + \left(\frac{y-ix}{2}\right)^2 + \left(\frac{u+iv}{2}\right)^2 + \left(\frac{v+iu}{2}\right)^2 = -2ixy + 2iuv$$

$$\implies xy - uv = 0 \tag{2.15}$$

and then modify this equation to the following matrix equation:

$$\begin{pmatrix} x & u \\ v & y \end{pmatrix} \begin{pmatrix} \eta_1 \\ \eta_2 \end{pmatrix} = 0 \tag{2.16}$$

for real parameters $\eta_1, \eta_2$.

The analysis to get the Kähler metric in these cases is more involved and is carried out in [5]. The results are limited cases of the following general form:

$$\begin{aligned} ds^2 &= \mathcal{G}_1\, dr^2 + \mathcal{G}_2 (d\psi + \cos\theta_1 d\phi_1 + \cos\theta_2 d\phi_2)^2 + \sum_{i=1}^{2} \mathcal{G}_{2+i}(d\theta_i^2 + \sin^2\theta_i d\phi_i^2) \\ &+ \mathcal{G}_5 \cos\psi\, (d\theta_1 d\theta_2 - \sin\theta_1 \sin\theta_2 d\phi_1 d\phi_2) + \mathcal{G}_6 \sin\psi\, (\sin\theta_1 d\phi_1 d\theta_2 + \sin\theta_2 d\phi_2 d\theta_1). \end{aligned} \tag{2.17}$$

In the Kähler deformed case, the warp factors $\mathcal{G}_i$ are given by (3.29), and in the Kähler resolved case the warp factors $\mathcal{G}_i$ are given by

$$\mathcal{G}_1 = \gamma, \quad \mathcal{G}_2 = \frac{\gamma r^2}{4}, \quad \mathcal{G}_3 = \frac{\gamma}{4}, \quad \mathcal{G}_4 = \frac{\gamma + 4a^2}{4}, \quad \mathcal{G}_5 = \mathcal{G}_6 = 0, \tag{2.18}$$

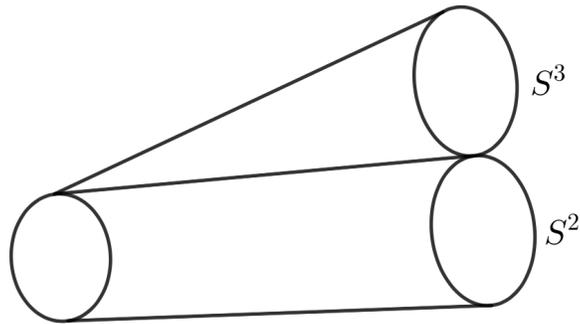

FIGURE 2.3: The resolved conifold

where $a$ is a constant known as the resolution parameter and $\gamma$ is a function of the Kähler potential[3]. As described in the two figures 2.2 and 2.3, the deformed conifold has its 2-sphere shrink to 0 radius at the tip, and the resolved conifold has its 3-sphere shrink to 0 radius at the tip.

When we take the $\mathcal{G}_i$ factors in (2.17) to become general, the metric then describes a complicated non-Kähler space with both resolution and deformation. It is our goal of the following chapter to determine which such metrics yields a dual $\mathcal{N} = 1$ SUSY gauge theory.

---

[3]The metric on a Kähler manifold satisfies $g_{\mu\bar{\nu}} = \partial_\mu \partial_{\bar{\nu}} \mathcal{F}$ for some function $\mathcal{F}$, known as the Kähler potential

# Chapter 3

# Calculating the supersymmetry constraints

We now have all of the background material to begin computing supersymmetry constraints. The calculation is long and tedious but it is straight forward. The main goal is to compute the three conditions listed at the end of 2.1 for the supergravity background that will be defined in (3.4). If those conditions are satisfied, then as stated in 2.1, our supergravity setup is dual to an $\mathcal{N} = 1$ supersymmetric gauge theory.

Before we begin we must distinguish between two different supergravity setups which are related to each other by a series of dualities. The details of the dualities themselves, described in [4], are not important to our calculation, but we will use the result to construct the supergravity background we wish to put SUSY constraints on.

The initial metric and flux configuration, which we will call the *pre-dual* setup, is given by

$$ds^2 = ds^2_{0123} + e^{-2\phi} ds^2_6, \qquad \mathbf{H} = e^{-2\phi} *_6 d\left(e^{2\phi} J\right), \tag{3.1}$$

where $\phi$ is the type IIB dilaton, and $J$ is the holomorphic $(1, 1)$ form constructed from this metric. Then as described in [4], after a series of 'duality chasing' is performed, we obtain the *post-dual* metric and flux configuration

$$ds^2 = \frac{1}{\sqrt{h}}\, ds^2_{0123} + \sqrt{h}\, ds^2_6,$$
$$\mathbf{F}_3 = \cosh\gamma e^{-2\phi} *_6 d\left(e^{2\phi} J\right), \qquad \mathbf{H}_3 = -\sinh\gamma\, d\left(e^{2\phi} J\right),$$
$$\widetilde{\mathbf{F}}_5 = -\sinh\gamma\cosh\gamma\,(1+*_{10})\,\mathcal{C}_5(r)\, d\psi \wedge \prod_{i=1}^{2} \sin\theta_i\, d\theta_i \wedge d\phi_i, \tag{3.2}$$

where $\gamma$ is a real number parameterizing the duality, $h$ is a function of $\gamma$, and $\mathcal{C}_5$ is a function of $r$. However, an important point to note is that the $J$ in (3.2) is the holomorphic $(1,1)$ form of the pre-dual setup (3.1); so we must compute $J$ in the pre-dual setup in order to construct the $\mathbf{H}_3$ and $\mathbf{F}_3$ fluxes in the post-dual setup.

This pre and post-dual setup was considered in [4], but with $ds_6^2$ being the metric for the non-Kähler resolved conifold

$$ds^2_{\text{resolved}} = \mathcal{G}_1\, dr^2 + \mathcal{G}_2(d\psi + \cos\theta_1 d\phi_1 + \cos\theta_2 d\phi_2)^2 + \sum_{i=1}^{2}\mathcal{G}_{2+i}(d\theta_i^2 + \sin^2\theta_i d\phi_i^2). \tag{3.3}$$

In our setup we will consider the more general case of a non-Kähler resolved deformed conifold with a metric given by

$$\begin{aligned}ds^2_6 &= \mathcal{G}_1\, dr^2 + \mathcal{G}_2(d\psi + \cos\theta_1 d\phi_1 + \cos\theta_2 d\phi_2)^2 + \sum_{i=1}^{2}\mathcal{G}_{2+i}(d\theta_i^2 + \sin^2\theta_i d\phi_i^2) \\ &+ \mathcal{G}_5 \cos\psi\,(d\theta_1 d\theta_2 - \sin\theta_1\sin\theta_2 d\phi_1 d\phi_2) + \mathcal{G}_6 \sin\psi\,(\sin\theta_1 d\phi_1 d\theta_2 + \sin\theta_2 d\phi_2 d\theta_1).\end{aligned} \tag{3.4}$$

The metric is *non-Kähler resolved*, because the factors $\mathcal{G}_i$, $i=1,\ldots,4$, both resolve the singular conifold and do not necessarily satisfy Kähler constraints, and it is *deformed* because of the inclusion of the $\mathcal{G}_5$ and $\mathcal{G}_6$ terms, which deforms the singular conifold as discussed in 2.1.

Because we can modify the geometry of the singular conifold in 3 ways—resolving, deforming, and warping—we view this as nearly the most general possible metric of a conifold. However we will make the simplification that the warp factors $\mathcal{G}_i$ are solely functions of $r$, and that $\mathcal{G}_5$ and $\mathcal{G}_6$ are equal:

$$\mathcal{G}_i = \mathcal{G}_i(r), \quad i = 1, \ldots, 6,$$

$$\mathcal{G}_5 = \mathcal{G}_6. \tag{3.5}$$

These assumptions drastically simplify calculations while still being more general than any case considered in the literature.

The general goal of the calculation is to generate the constraints the warp factors $\mathcal{G}_i$ must satisfy in order for $\mathcal{N} = 1$ supersymmetry to hold.

## 3.1 Deriving a set of vielbeins for the internal space

Many parts of the calculation are simplified if we write the metric $ds_6^2$ in terms of vielbeins. That is, we find a basis $e_i$ satisfying

$$ds_6^2 = \sum_{i=1}^{6} e_i^2, \tag{3.6}$$

where $e_i^2 \equiv e_i \otimes e_i$.

We start by defining the left invariant Maurer-Cartan forms

$$\begin{pmatrix} \sigma_1 \\ \sigma_2 \end{pmatrix} = \begin{pmatrix} \cos \psi_1 & \sin \psi_1 \\ -\sin \psi_1 & \cos \psi_1 \end{pmatrix} \begin{pmatrix} d\theta_1 \\ \sin \theta_1 \, d\phi_1 \end{pmatrix} = \begin{pmatrix} \cos \psi_1 \, d\theta_1 + \sin \psi_1 \sin \theta_1 \, d\phi_1 \\ -\sin \psi_1 \, d\theta_1 + \cos \psi_1 \sin \theta_1 \, d\phi_1 \end{pmatrix}$$

$$\begin{pmatrix} \Sigma_1 \\ \Sigma_2 \end{pmatrix} = \begin{pmatrix} \cos\psi_2 & \sin\psi_2 \\ -\sin\psi_2 & \cos\psi_2 \end{pmatrix} \begin{pmatrix} d\theta_2 \\ \sin\theta_2\, d\phi_2 \end{pmatrix} = \begin{pmatrix} \cos\psi_2\, d\theta_2 + \sin\psi_2 \sin\theta_2\, d\phi_2 \\ -\sin\psi_2\, d\theta_2 + \cos\psi_2 \sin\theta_2\, d\phi_2 \end{pmatrix}$$

$$\sigma_3 = d\psi_1 + \cos\theta_1\, d\phi_1, \quad \Sigma_3 = d\psi_2 + \cos\theta_2\, d\phi_2. \tag{3.7}$$

Using these forms we can construct an ansatz for the vielbeins that describe the internal part of the metric $\sqrt{h}ds_6^2$ (3.4). We will simplify the analysis by taking $h=1$; this does not change any results from our analysis since we never take derivatives of any quantities dependent on the post-dual vielbeins. The correct vielbeins, however, will contain a factor of $\sqrt{\sqrt{h}}$ so that their sum squared yields the $\sqrt{h}$ factor in the metric, and this factor can be placed back in afterwards. We write an ansatz for the vielbeins as the following linear combinations of the left invariant Maurer-Cartan forms (3.7) (besides $e_1$ which is just proportional to $dr$):

$$\begin{aligned}
e_1 &= \sqrt{\mathcal{G}_1}\, dr, \\
e_2 &= \sqrt{\mathcal{G}_2}\, [\sigma_3 + \Sigma_3] = \sqrt{\mathcal{G}_2}\, [d\psi_1 + d\psi_2 + d\phi_1 \cos\theta_1 + d\phi_2 \cos\theta_2] \\
e_3 &= \sqrt{\mathcal{G}_3}\, [\alpha_1 \sigma_1 + \beta_3 \Sigma_1] \\
&= \sqrt{\mathcal{G}_3}\, [\alpha_1 (d\theta_1 \cos\psi_1 + d\phi_1 \sin\theta_1 \sin\psi_1) + \beta_3 (d\theta_2 \cos\psi_2 + d\phi_2 \sin\theta_2 \sin\psi_2)] \\
e_4 &= \sqrt{\mathcal{G}_4}\, [\alpha_2 \sigma_2 - \beta_4 \Sigma_2] \\
&= \sqrt{\mathcal{G}_4}\, [\alpha_2 (d\phi_1 \sin\theta_1 \cos\psi_1 - d\theta_1 \sin\psi_1) - \beta_4 (d\phi_2 \sin\theta_2 \cos\psi_2 - d\theta_2 \sin\psi_2)] \\
e_5 &= \sqrt{\mathcal{G}_3}\, [\beta_1 \sigma_1 + \alpha_3 \Sigma_1] \\
&= \sqrt{\mathcal{G}_3}\, [\alpha_3 (d\theta_2 \cos\psi_2 + d\phi_2 \sin\theta_2 \sin\psi_2) + \beta_1 (d\theta_1 \cos\psi_1 + d\phi_1 \sin\theta_1 \sin\psi_1)] \\
e_6 &= \sqrt{\mathcal{G}_4}\, [-\beta_2 \sigma_2 + \alpha_4 \Sigma_2] \\
&= \sqrt{\mathcal{G}_4}\, [\alpha_4 (d\phi_2 \sin\theta_2 \cos\psi_2 - d\theta_2 \sin\psi_2) - \beta_2 (d\phi_1 \sin\theta_1 \cos\psi_1 - d\theta_1 \sin\psi_1)]
\end{aligned} \tag{3.8}$$

We have defined these vielbeins using 8 undetermined parameters $\alpha_i$, $\beta_i$, $i = 1, \ldots, 4$ which will be restricted shortly. Also we have an extra coordinate redundancy because of the appearance of the angular variables $\psi_1$ and $\psi_2$, instead of just having the angular variable $\psi$ as in (3.4). The $\psi_1$ and $\psi_2$ variables are related to $\psi$ by the equation

$$\psi = \psi_1 + \psi_2, \tag{3.9}$$

and we will eventually restrict to the subspace $\psi_1 = \psi_2 = \psi/2$ to eliminate the extra degree of freedom in our definition of the vielbeins. Summing the squares of these vielbeins leads to the expression

$$\begin{aligned}ds_6^2 &= \mathcal{G}_1 \, dr^2 + \mathcal{G}_2 (d\psi + \cos\theta_1 d\phi_1 + \cos\theta_2 d\phi_2)^2 \\ &+ \mathcal{G}_3 \left[ d\phi_1^2 \sin^2\theta_1 \left( \left(\alpha_1^2 + \beta_1^2\right) \sin^2\psi_1 + \frac{\mathcal{G}_4}{\mathcal{G}_3} \left(\alpha_2^2 + \beta_2^2\right) \cos^2\psi_1 \right) \right. \\ &\left. + d\theta_1^2 \left( \frac{\mathcal{G}_4}{\mathcal{G}_3} \left(\alpha_2^2 + \beta_2^2\right) \sin^2\psi_1 + \left(\alpha_1^2 + \beta_1^2\right) \cos^2\psi_1 \right) \right] \\ &+ \mathcal{G}_4 \left[ d\phi_2^2 \sin^2\theta_2 \left( \frac{\mathcal{G}_3}{\mathcal{G}_4} \left(\alpha_3^2 + \beta_3^2\right) \sin^2\psi_2 + \left(\alpha_4^2 + \beta_4^2\right) \cos^2\psi_2 \right) \right. \\ &\left. + d\theta_2^2 \left( \left(\alpha_4^2 + \beta_4^2\right) \sin^2\psi_2 + \frac{\mathcal{G}_3}{\mathcal{G}_4} \left(\alpha_3^2 + \beta_3^2\right) \cos^2\psi_2 \right) \right] \\ &+ \mathcal{G}_6 \left[ 2d\theta_1 d\theta_2 \left( \frac{\mathcal{G}_3}{\mathcal{G}_6} (\alpha_3\beta_1 + \alpha_1\beta_3) \cos\psi_1 \cos\psi_2 - \frac{\mathcal{G}_4}{\mathcal{G}_6} (\alpha_4\beta_2 + \alpha_2\beta_4) \sin\psi_1 \sin\psi_2 \right) \right. \quad (3.10) \\ &\left. + 2d\phi_1 d\phi_2 \sin\theta_1 \sin\theta_2 \left( \frac{\mathcal{G}_3}{\mathcal{G}_6} (\alpha_3\beta_1 + \alpha_1\beta_3) \sin\psi_1 \sin\psi_2 - \frac{\mathcal{G}_4}{\mathcal{G}_6} (\alpha_4\beta_2 + \alpha_2\beta_4) \cos\psi_1 \cos\psi_2 \right) \right] \\ &+ \mathcal{G}_6 \left[ d\theta_1 d\phi_2 \sin\theta_2 \left( \frac{\mathcal{G}_3}{\mathcal{G}_6} (\alpha_3\beta_1 + \alpha_1\beta_3) \sin\psi_2 \cos\psi_1 + \frac{\mathcal{G}_4}{\mathcal{G}_6} (\alpha_4\beta_2 + \alpha_2\beta_4) \sin\psi_1 \cos\psi_2 \right) \right. \\ &\left. + 2d\theta_2 d\phi_1 \sin\theta_1 \left( \frac{\mathcal{G}_3}{\mathcal{G}_6} (\alpha_3\beta_1 + \alpha_1\beta_3) \sin\psi_1 \cos\psi_2 + \frac{\mathcal{G}_4}{\mathcal{G}_6} (\alpha_4\beta_2 + \alpha_2\beta_4) \sin\psi_2 \cos\psi_1 \right) \right]. \end{aligned}$$

$$\tag{3.11}$$

We want this expression to match our expression for $ds_6^2$ in (3.4). Right away we see that the $\mathcal{G}_1$ and $\mathcal{G}_2$ terms match. For the rest of the terms, however, we need to force constraints on the $\alpha_i$ and $\beta_i$ in order to recover the correct expression. For the $\mathcal{G}_3$ terms we require the terms multiplying $d\phi_1^2\sin^2\theta_1$ and $d\theta_1^2$ to be equal to 1. Using the trigonometric identity $\sin^2\psi_1 + \cos^2\psi_1 = 1$ leads us to the following constraints:

$$\alpha_1^2 + \beta_1^2 = 1, \quad \alpha_2^2 + \beta_2^2 = \frac{\mathcal{G}_3}{\mathcal{G}_4}. \tag{3.12}$$

When those constraints hold, the $\psi_1$ dependent terms cancel in both right brackets and we are left with the correct $\mathcal{G}_3$ term as in (3.4). Similarly for the $\mathcal{G}_4$ term we require

$$\alpha_4^2 + \beta_4^2 = 1, \quad \alpha_3^2 + \beta_3^2 = \frac{\mathcal{G}_4}{\mathcal{G}_3}. \tag{3.13}$$

Notice we have two $\mathcal{G}_6$ terms; this is due to setting $\mathcal{G}_5 = \mathcal{G}_6$, which is one of the assumptions of our model that we made in (3.5). For the first $\mathcal{G}_6$ term we want to combine the $\cos\psi_i$ and $\sin\psi_i$ terms into a $\cos\psi$ for the term to match. To do this we use the trigonometric identity $\cos(\psi_1 + \psi_2) = \cos\psi_1\cos\psi_2 - \sin\psi_1\sin\psi_2$. This leads to the constraints

$$\alpha_3\beta_1 + \alpha_1\beta_3 = \frac{\mathcal{G}_6}{2\mathcal{G}_3}, \quad \alpha_4\beta_2 + \alpha_2\beta_4 = \frac{\mathcal{G}_6}{2\mathcal{G}_4}, \tag{3.14}$$

where the factor of $\frac{1}{2}$ is there to cancel the factor of 2 in front of the $d\theta_1 d\theta_2$ term. With those substitutions, the first and second $\mathcal{G}_6$ terms automatically match with those in (3.4) and we are done. These 6 equations along with (3.8) define the vielbeins we will be using for our general metric.

As a consistency check, we note that for the following values of $\alpha_i$ and $\beta_i$ the vielbeins reduce to those of the non-Kähler resolved case (3.3):

$$\alpha_1 = \alpha_4 = 1, \quad \alpha_2 = \sqrt{\frac{\mathcal{G}_3}{\mathcal{G}_4}}, \quad \alpha_3 = \sqrt{\frac{\mathcal{G}_4}{\mathcal{G}_3}}, \quad \beta_i = 0. \tag{3.15}$$

This can be seen by plugging in these values into the expression (3.10).

## 3.2 Computing $J$ and $\Omega$ in the pre-dual setup

As explained in 2.1, a 6-dimensional complex manifold with SU(3) structure is mainly characterized by its fundamental $(1,1)$ form $J$ and its holomorphic $(3,0)$ form $\Omega$. In order for the post-dual setup to be describe a supergravity theory that is $\mathcal{N} = 1$ supersymmetric, we require the pre-dual theory to live on a complex manifold. From 2.1, we know that a manifold is complex if its holomorphic $(3,0)$ form is closed:

$$d\Omega = 0. \tag{3.16}$$

In this section we will show that it is possible to have complex manifold with our generalized metric (3.4).

We also need to compute the fundamental $(1,1)$ form $J$ for two reasons. The first reason is to determine whether the pre-dual manifold under consideration is Kähler; it is Kähler in the case where $d(e^{2\phi}J) = 0$ and it is non-Kähler otherwise. The second reason is to be able to compute the 3-form fluxes $\mathbf{H}_3$ and $\mathbf{F}_3$ of the post-dual setup. These quantities are important because we need them to compute the complex 3-form flux $\mathbf{G}_3 = \mathbf{F}_3 - ie^{-\phi}\mathbf{H}_3$, which is pivotal in determining the SUSY constraints.

To calculate these quantities, we will need a basis of complex vielbeins $\mathcal{E}_i$, which we define as

$$\begin{aligned}
\mathcal{E}_1 &= e^{-\phi}(e_2 + ie_1) = e^{-\phi}\left[i\sqrt{\mathcal{G}_1}dr + \sqrt{\mathcal{G}_2}\left(d\psi + \cos\theta_1 d\phi_1 + \cos\theta_2 d\phi_2\right)\right] \\
\mathcal{E}_2 &= e^{-\phi}(e_3 + ie_4) = e^{-\phi}\left[\sqrt{\mathcal{G}_3}\left(\alpha_1\sigma_1 + \beta_3\Sigma_1\right) + i\sqrt{\mathcal{G}_4}\left(\alpha_2\sigma_2 - \beta_4\Sigma_2\right)\right] \\
\mathcal{E}_3 &= e^{-\phi}(e_5 + ie_6) = e^{-\phi}\left[\sqrt{\mathcal{G}_3}\left(\beta_1\sigma_1 + \alpha_3\Sigma_1\right) - i\sqrt{\mathcal{G}_4}\left(\beta_2\sigma_2 - \alpha_4\Sigma_2\right)\right],
\end{aligned} \tag{3.17}$$

where $e^{-2\phi}ds_6^2 = \mathcal{E}_1 \otimes \overline{\mathcal{E}}_1 + \mathcal{E}_2 \otimes \overline{\mathcal{E}}_2 + \mathcal{E}_3 \otimes \overline{\mathcal{E}}_3$. Notice we have accounted for the warping $e^{-2\phi}ds_6^2$ of the pre-dual setup (3.1) unlike before where we set the post-dual warp factor $h = 1$; this is because the analysis in this section will involve taking derivatives of the complex vielbeins so it is crucial to include it. The fundamental $(1,1)$ form is calculated as in 2.1:

$$\begin{aligned} J &= -\frac{i}{2}\left(\mathcal{E}_1 \wedge \overline{\mathcal{E}}_1 + \mathcal{E}_2 \wedge \overline{\mathcal{E}}_2 + \mathcal{E}_3 \wedge \overline{\mathcal{E}}_3\right) \\ &= e^{-2\phi}\sqrt{\mathcal{G}_1\mathcal{G}_2}\, dr \wedge (d\psi + \cos\theta_1 d\phi_1 + \cos\theta_2 d\phi_2) \\ &\quad - e^{-2\phi}\sqrt{\mathcal{G}_3\mathcal{G}_4}\left[(\alpha_1\alpha_2 - \beta_1\beta_2)\,\sigma_1 \wedge \sigma_2 + (\alpha_3\alpha_4 - \beta_3\beta_4)\,\Sigma_1 \wedge \Sigma_2\right] \\ &\quad - e^{-2\phi}\sqrt{\mathcal{G}_3\mathcal{G}_4}\left[(\beta_1\alpha_4 - \alpha_1\beta_4)\,\sigma_1 \wedge \Sigma_2 + (\beta_3\alpha_2 - \alpha_3\beta_2)\,\Sigma_1 \wedge \sigma_2\right]. \end{aligned} \quad (3.18)$$

We will define a set of basis vectors that will help simplify the appearance of many expressions in the rest of the calculation:

$$\begin{aligned} e_r &= dr, \quad e_\psi = d\psi + \cos\theta_1 d\phi_1 + \cos\theta_2 d\phi_2 \\ e_{\phi_1} &= \sin\theta_1 d\phi_1, \quad e_{\phi_2} = \sin\theta_2 d\phi_2 \\ e_{\theta_1} &= d\theta_1, \quad e_{\theta_2} = d\theta_2. \end{aligned} \quad (3.19)$$

Using these basis vectors, we can rewrite the $\sigma_i$ and $\Sigma_j$ terms as

$$\begin{aligned} \sigma_1 &= \cos\frac{\psi}{2}e_{\theta_1} + \sin\frac{\psi}{2}e_{\phi_1}, \quad \sigma_2 = -\sin\frac{\psi}{2}e_{\theta_1} + \cos\frac{\psi}{2}e_{\phi_1} \\ \Sigma_1 &= \cos\frac{\psi}{2}e_{\theta_2} + \sin\frac{\psi}{2}e_{\phi_2}, \quad \Sigma_2 = -\sin\frac{\psi}{2}e_{\theta_2} + \cos\frac{\psi}{2}e_{\phi_2}, \end{aligned} \quad (3.20)$$

where we have set $\psi_1 = \psi_2 = \frac{\psi}{2}$ to eliminate the extra coordinate redundancy discussed earlier. We also define the coefficients

$$\mathbf{A} = \beta_1\alpha_4 - \alpha_1\beta_4, \quad \mathbf{B} = \beta_3\alpha_2 - \alpha_3\beta_2, \quad (3.21)$$

to condense the fourth line of (3.18). Using (3.20) and (3.21), we rewrite the fundamental form $J$ as

$$\begin{aligned}
e^{2\phi} J &= \sqrt{\mathcal{G}_1 \mathcal{G}_2} dr \wedge (d\psi + \cos\theta_1 d\phi_1 + \cos\theta_2 d\phi_2) \\
&\quad - \frac{1}{2}\sqrt{\mathcal{G}_3 \mathcal{G}_4} \left(\mathbf{A} - \mathbf{B}\right) \sin\psi \left(e_{\phi_1} \wedge e_{\phi_2} - e_{\theta_1} \wedge e_{\theta_2}\right) \\
&\quad - \sqrt{\mathcal{G}_3 \mathcal{G}_4} \left[(\alpha_1 \alpha_2 - \beta_1 \beta_2) e_{\theta_1} \wedge e_{\phi_1} + (\alpha_3 \alpha_4 - \beta_3 \beta_4) e_{\theta_2} \wedge e_{\phi_2}\right] \\
&\quad - \sqrt{\mathcal{G}_3 \mathcal{G}_4} \left[\left(\mathbf{A} \cos^2 \frac{\psi}{2} + \mathbf{B} \sin^2 \frac{\psi}{2}\right) e_{\theta_1} \wedge e_{\phi_2} + \left(\mathbf{B} \cos^2 \frac{\psi}{2} + \mathbf{A} \sin^2 \frac{\psi}{2}\right) e_{\theta_2} \wedge e_{\phi_1}\right].
\end{aligned} \tag{3.22}$$

At this point we can explicitly show that this metric is non-Kähler for certain parameters. We will compute the closure of $J$ in the case

$$\mathbf{A} = \mathbf{B} \tag{3.23}$$

and show it is non-zero. It turns out, as we will show in 3.4, that once we impose that $\mathbf{G}_3$ be an ISD $(2,1)$ form, one of the conditions that appears is $\mathbf{A} = \mathbf{B} = 0$, so (3.23) is not actually a simplification.

We now define the following functions that will simplify the expression in this limit:

$$\begin{aligned}
\mathbf{C} &\equiv \sqrt{\mathcal{G}_3 \mathcal{G}_4}(\alpha_1 \alpha_2 - \beta_1 \beta_2), \quad \mathbf{D} \equiv \sqrt{\mathcal{G}_3 \mathcal{G}_4}(\alpha_3 \alpha_4 - \beta_3 \beta_4) \\
F &\equiv \mathbf{A}\sqrt{\mathcal{G}_3 \mathcal{G}_4}.
\end{aligned} \tag{3.24}$$

Setting $\mathbf{A} = \mathbf{B}$, both the second line and the $\psi$ dependence of the fourth line of (3.22) vanishes:

$$\begin{aligned}
e^{2\phi} J &= \sqrt{\mathcal{G}_1 \mathcal{G}_2} e_r \wedge e_\psi \\
&\quad - \mathbf{C} e_{\theta_1} \wedge e_{\phi_1} - \mathbf{D} e_{\theta_2} \wedge e_{\phi_2} \\
&\quad - F \left[e_{\theta_1} \wedge e_{\phi_2} + e_{\theta_2} \wedge e_{\phi_1}\right].
\end{aligned} \tag{3.25}$$

To compute $d(e^{2\phi}J)$, we will make frequent use of the following identities:

$$\begin{aligned}
d(e_r) &= d^2 r = 0 \\
d(e_\psi) &= d(d\psi + \cos\theta_1 d\phi_1 + \cos\theta_2 d\phi_2) = -\sin\theta_1 d\theta_1 \wedge d\phi_1 - \sin\theta_2 d\theta_2 \wedge d\phi_2 \\
&= -(e_{\theta_1} \wedge e_{\phi_1} + e_{\theta_2} \wedge e_{\phi_2}) \\
d(e_{\theta_1}) &= d(d\theta_1) = 0, \quad d(e_{\theta_2}) = d(d\theta_2) = 0 \\
d(e_{\phi_1}) &= d(\sin\theta_1 d\phi_1) = \cos\theta_1 d\theta_1 \wedge d\phi_1 = \cot\theta_1 e_{\theta_1} \wedge e_{\phi_1} \\
d(e_{\phi_2}) &= d(\sin\theta_2 d\phi_2) = \cos\theta_2 d\theta_2 \wedge d\phi_2 = \cot\theta_2 e_{\theta_2} \wedge e_{\phi_2}.
\end{aligned} \quad (3.26)$$

Using these we have

$$\begin{aligned}
d(e^{2\phi}J) &= \sqrt{\mathcal{G}_1 \mathcal{G}_2} e_r \wedge (e_{\theta_1} \wedge e_{\phi_1} + e_{\theta_2} \wedge e_{\phi_2}) \\
&\quad - \mathbf{C}_r e_r \wedge e_{\theta_1} \wedge e_{\phi_1} - \mathbf{D}_r e_r \wedge e_{\theta_2} \wedge e_{\phi_2} \\
&\quad - F_r e_r \wedge (e_{\theta_1} \wedge e_{\phi_2} + e_{\theta_2} \wedge e_{\phi_1}) + F\left[e_{\theta_1} \wedge (\cot\theta_2 e_{\theta_2} \wedge e_{\phi_2}) + e_{\theta_2} \wedge (\cot\theta_1 e_{\theta_1} \wedge e_{\phi_1})\right] \\
&= dr \wedge \left[e_{\theta_1} \wedge e_{\phi_1}\left(\sqrt{\mathcal{G}_1\mathcal{G}_2} - \frac{\partial \mathbf{C}}{\partial r}\right) + e_{\theta_2} \wedge e_{\phi_2}\left(\sqrt{\mathcal{G}_1\mathcal{G}_2} - \frac{\partial \mathbf{D}}{\partial r}\right)\right] \\
&\quad - \frac{\partial \mathcal{F}}{\partial r} dr \wedge (e_{\theta_1} \wedge e_{\phi_2} + e_{\theta_2} \wedge e_{\phi_1}) - \mathcal{F} \, d\theta_1 \wedge d\theta_2 \wedge (\cos\theta_1 d\phi_1 - \cos\theta_2 d\phi_2), \quad (3.27)
\end{aligned}$$

which is generically non-zero which implies that the manifold is non-Kähler. However the manifold is Kähler when all of the coefficients in the expression vanish, namely when

$$\begin{aligned}
\frac{\partial \mathcal{F}}{\partial r} &= \mathcal{F} = 0 \\
\sqrt{\mathcal{G}_1 \mathcal{G}_2} &= \frac{\partial \mathbf{C}}{\partial r} = \frac{\partial \mathbf{D}}{\partial r}.
\end{aligned} \quad (3.28)$$

We will test out these constraints on a known Kähler metric, the Kähler deformed conifold[1]:

$$\mathcal{G}_3 = \mathcal{G}_4 = \frac{\gamma}{4}, \quad \mathcal{G}_6 = \frac{\mu^2 \gamma}{2r^2}$$

$$\mathcal{G}_1 = \frac{\gamma + (r^2\gamma' - \gamma)\left(1 - \frac{\mu^4}{r^4}\right)}{r^2\left(1 - \frac{\mu^4}{r^4}\right)}, \quad \mathcal{G}_2 = \frac{1}{4}\left[\gamma + (r^2\gamma' - \gamma)\left(1 - \frac{\mu^4}{r^4}\right)\right], \quad (3.29)$$

where $\mu^2$ is a function of the deformation parameter and $\gamma$ (unrelated to the one we defined in (3.2)) is a function of the Kähler potential $\gamma = r^2 \mathbf{F}$, and $\gamma' \equiv \frac{d}{d(r^2)}\gamma$. As shown in [5], this leads to the following coefficients for the vielbeins:

$$\alpha_i \equiv \alpha = \frac{1}{2}\sqrt{1 + \frac{\mu^2}{r^2}} + \frac{1}{2}\sqrt{1 - \frac{\mu^2}{r^2}}, \quad \beta_i \equiv \beta = \frac{\mu^2}{2r^2\alpha}. \quad (3.30)$$

Plugging these into (3.24), $\mathbf{C}, \mathbf{D}$, and $\mathcal{F}$ become

$$\mathbf{C} = \mathbf{D} = \frac{1}{4}\gamma\left(1 - \frac{\mu^4}{r^4\left(\sqrt{1 - \frac{\mu^4}{r^4}} + 1\right)}\right), \quad \mathcal{F} = 0. \quad (3.31)$$

Since $\mathcal{F} = 0$, the first equation of (3.28) is satisfied. Using the chain rule $\frac{\partial \mathbf{C}}{\partial r} = \frac{\partial \mathbf{C}}{\partial r^2}\frac{dr^2}{dr} = 2r\frac{\partial \mathbf{C}}{\partial r^2}$, we have that

$$\frac{\partial \mathbf{C}}{\partial r} = \frac{2\gamma(\mu^4 + r^2(r^4 - \mu^4))}{r^3}\sqrt{1 - \frac{\mu^4}{r^4}}. \quad (3.32)$$

Plugging in (3.32) and (3.29) into (3.28) yields the differential equation

$$r^2\left(r^4 - \mu^4\right)\frac{d\gamma^3}{dr^2} + 3r^4\gamma^3 = 2r^8, \quad (3.33)$$

which exactly matches the differential equation for $\gamma$ derived in [2].

---
[1]We use the metric calculated in [5]

We now wish to determine whether the manifold in the pre-dual setup is complex. To do this we will compute the closure of the holomorphic $(3,0)$ form $\Omega$, which is defined as

$$\Omega \equiv e^{3\phi}\mathcal{E}_1 \wedge \mathcal{E}_2 \wedge \mathcal{E}_3$$

$$= \left[i\sqrt{\mathcal{G}_1}e_r + \sqrt{\mathcal{G}_2}e_\psi\right] \wedge \left[\sqrt{\mathcal{G}_3}\left(\alpha_1\sigma_1 + \beta_3\Sigma_1\right) + i\sqrt{\mathcal{G}_4}\left(\alpha_2\sigma_2 - \beta_4\Sigma_2\right)\right]$$

$$\wedge \left[\sqrt{\mathcal{G}_3}\left(\beta_1\sigma_1 + \alpha_3\Sigma_1\right) - i\sqrt{\mathcal{G}_4}\left(\beta_2\sigma_2 - \alpha_4\Sigma_2\right)\right]$$

$$= \left[i\sqrt{\mathcal{G}_1}e_r + \sqrt{\mathcal{G}_2}e_\psi\right] \wedge \left[\sqrt{\mathcal{G}_3}\left(\alpha_1(\cos\frac{\psi}{2}e_{\theta_1} + \sin\frac{\psi}{2}e_{\phi_1}) + \beta_3(\cos\frac{\psi}{2}e_{\theta_2} + \sin\frac{\psi}{2}e_{\phi_2})\right)\right.$$

$$\left. + i\sqrt{\mathcal{G}_4}\left(\alpha_2(-\sin\frac{\psi}{2}e_{\theta_1} + \cos\frac{\psi}{2}e_{\phi_1}) - \beta_4(-\sin\frac{\psi}{2}e_{\theta_2} + \cos\frac{\psi}{2}e_{\phi_2})\right)\right] \wedge$$

$$\left[\sqrt{\mathcal{G}_3}\left(\beta_1(\cos\frac{\psi}{2}e_{\theta_1} + \sin\frac{\psi}{2}e_{\phi_1}) + \alpha_3(\cos\frac{\psi}{2}e_{\theta_2} + \sin\frac{\psi}{2}e_{\phi_2})\right)\right.$$

$$\left. - i\sqrt{\mathcal{G}_4}\left(\beta_2(-\sin\frac{\psi}{2}e_{\theta_1} + \cos\frac{\psi}{2}e_{\phi_1})\alpha_4(-\sin\frac{\psi}{2}e_{\theta_2} + \cos\frac{\psi}{2}e_{\phi_2})\right)\right].$$

Expanding the expression and collecting terms, $\Omega$ becomes

$$\begin{aligned}\Omega &= e_r \wedge (\mathbf{A}_{11}\ e_{\theta_2} \wedge e_{\phi_2} - \mathbf{A}_{12}\ e_{\theta_1} \wedge e_{\phi_1}) + e_r \wedge (\mathbf{A}_{21}\ e_{\phi_1} \wedge e_{\theta_2} + \mathbf{A}_{22}\ e_{\theta_1} \wedge e_{\phi_2}) \\ &+ e_r \wedge (\mathbf{A}_{31}\ e_{\theta_1} \wedge e_{\theta_2} - \mathbf{A}_{32}\ e_{\phi_1} \wedge e_{\phi_2}) + e_\psi \wedge (\mathbf{A}_{41}\ e_{\theta_2} \wedge e_{\phi_2} - \mathbf{A}_{42}\ e_{\theta_1} \wedge e_{\phi_1}) \\ &+ e_\psi \wedge (\mathbf{A}_{51}\ e_{\phi_1} \wedge e_{\theta_2} + \mathbf{A}_{52}\ e_{\theta_1} \wedge e_{\phi_2}) + e_\psi \wedge (\mathbf{A}_{61}\ e_{\theta_1} \wedge e_{\theta_2} - \mathbf{A}_{62}\ e_{\phi_1} \wedge e_{\phi_2}),\end{aligned}$$
(3.34)

where the $\mathbf{A}_{ij}$ terms are defined as follows:

$$\begin{aligned}\mathbf{A}_{11} &= -\sqrt{\mathcal{G}_1\mathcal{G}_3\mathcal{G}_4}\left(\alpha_4\beta_3 + \alpha_3\beta_4\right), \qquad \mathbf{A}_{12} = -\sqrt{\mathcal{G}_1\mathcal{G}_3\mathcal{G}_4}\left(\alpha_1\beta_2 + \alpha_2\beta_1\right) \\ \mathbf{A}_{21} &= i\sin\frac{\psi}{2}\cos\frac{\psi}{2}\mathcal{G}_3\sqrt{\mathcal{G}_1}\left(\alpha_1\alpha_3 - \beta_1\beta_3\right) + i\cos\frac{\psi}{2}\sin\frac{\psi}{2}\mathcal{G}_4\sqrt{\mathcal{G}_1}\left(\alpha_2\alpha_4 - \beta_2\beta_4\right) \\ &+ \sin^2\frac{\psi}{2}\sqrt{\mathcal{G}_1\mathcal{G}_3\mathcal{G}_4}\left(\alpha_1\alpha_4 + \beta_1\beta_4\right) - \cos^2\frac{\psi}{2}\sqrt{\mathcal{G}_1\mathcal{G}_3\mathcal{G}_4}\left(\alpha_2\alpha_3 + \beta_2\beta_3\right) \\ \mathbf{A}_{22} &= i\cos\frac{\psi}{2}\sin\frac{\psi}{2}\mathcal{G}_3\sqrt{\mathcal{G}_1}\left(\alpha_1\alpha_3 - \beta_1\beta_3\right) + i\sin\frac{\psi}{2}\cos\frac{\psi}{2}\mathcal{G}_4\sqrt{\mathcal{G}_1}\left(\alpha_2\alpha_4 - \beta_2\beta_4\right) \\ &- \cos^2\frac{\psi}{2}\sqrt{\mathcal{G}_1\mathcal{G}_3\mathcal{G}_4}\left(\alpha_1\alpha_4 + \beta_1\beta_4\right) + \sin^2\frac{\psi}{2}\sqrt{\mathcal{G}_1\mathcal{G}_3\mathcal{G}_4}\left(\alpha_2\alpha_3 + \beta_2\beta_3\right)\end{aligned}$$

$$\begin{aligned}
\mathbf{A}_{31} &= i\cos^2\frac{\psi}{2}\,\mathcal{G}_3\sqrt{\mathcal{G}_1}\,(\alpha_1\alpha_3 - \beta_1\beta_3) - i\sin^2\frac{\psi}{2}\,\mathcal{G}_4\sqrt{\mathcal{G}_1}\,(\alpha_2\alpha_4 - \beta_2\beta_4) \\
&+ \cos\frac{\psi}{2}\sin\frac{\psi}{2}\,\sqrt{\mathcal{G}_1\mathcal{G}_3\mathcal{G}_4}\,(\alpha_1\alpha_4 + \beta_1\beta_4) + \cos\frac{\psi}{2}\sin\frac{\psi}{2}\,\sqrt{\mathcal{G}_1\mathcal{G}_3\mathcal{G}_4}\,(\alpha_2\alpha_3 + \beta_2\beta_3) \\
\mathbf{A}_{32} &= -i\sin^2\frac{\psi}{2}\,\mathcal{G}_3\sqrt{\mathcal{G}_1}\,(\alpha_1\alpha_3 - \beta_1\beta_3) + i\cos^2\frac{\psi}{2}\,\mathcal{G}_4\sqrt{\mathcal{G}_1}\,(\alpha_2\alpha_4 - \beta_2\beta_4) \\
&+ \sin\frac{\psi}{2}\cos\frac{\psi}{2}\,\sqrt{\mathcal{G}_1\mathcal{G}_3\mathcal{G}_4}\,(\alpha_1\alpha_4 + \beta_1\beta_4) + \cos\frac{\psi}{2}\sin\frac{\psi}{2}\,\sqrt{\mathcal{G}_1\mathcal{G}_3\mathcal{G}_4}\,(\alpha_2\alpha_3 + \beta_2\beta_3), \\
\mathbf{A}_{nk} &\equiv -i\,\mathbf{A}_{n-3,k}\sqrt{\frac{\mathcal{G}_2}{\mathcal{G}_1}}, \quad n \geq 4.
\end{aligned}$$
(3.35)

We will begin by verifying the closure of $\Omega$ in the simpler Kähler deformed case defined in (3.29) and (3.30). In this limit, the $\mathbf{A}_{ij}$ coefficients simplify to

$$\begin{aligned}
\mathbf{A}_{11} &= \mathbf{A}_{12} = -2\alpha\beta\mathcal{G}_3\sqrt{\mathcal{G}_1} \\
\mathbf{A}_{21} &= \mathbf{A}_{22} = i\mathcal{G}_3\sqrt{\mathcal{G}_1}((\alpha^2 - \beta^2)\sin\psi + i(\alpha^2+\beta^2)\cos\psi) = i\mathcal{G}_3\sqrt{\mathcal{G}_1}((\alpha^2-\beta^2)\sin\psi + i\cos\psi) \\
\mathbf{A}_{31} &= \mathbf{A}_{32} = i\mathcal{G}_3\sqrt{\mathcal{G}_1}((\alpha^2 - \beta^2)\cos\psi - i(\alpha^2+\beta^2)\sin\psi) = i\mathcal{G}_3\sqrt{\mathcal{G}_1}((\alpha^2-\beta^2)\cos\psi - i\sin\psi) \\
\mathbf{A}_{nk} &= -i\,\mathbf{A}_{n-3,k}\sqrt{\frac{\mathcal{G}_2}{\mathcal{G}_1}}, \quad n \geq 4,
\end{aligned}$$

where we have used the trigonometric identities $2\sin\frac{\psi}{2}\cos\frac{\psi}{2} = \sin\psi$ and $\cos^2\frac{\psi}{2} - \sin^2\frac{\psi}{2} = \cos\psi$, and the constraint $\alpha^2 + \beta^2 = 1$. $\Omega$ then simplifies to

$$\begin{aligned}
\Omega &= 2i\,\mathcal{G}_3\,\alpha\beta\,(e_{\theta_2}\wedge e_{\phi_2} - e_{\theta_1}\wedge e_{\phi_1})\wedge\left(i\sqrt{\mathcal{G}_1}\,e_r + \sqrt{\mathcal{G}_2}\,e_\psi\right) \\
&+ i\left[(\alpha^2 - \beta^2)\sin\psi + i\cos\psi\right]\mathcal{G}_3\sqrt{\mathcal{G}_1}\,e_r\wedge(e_{\phi_1}\wedge e_{\theta_2} + e_{\theta_1}\wedge e_{\phi_2}) \\
&+ i\left[(\alpha^2 - \beta^2)\cos\psi - i\sin\psi\right]\mathcal{G}_3\sqrt{\mathcal{G}_1}\,e_r\wedge(e_{\theta_1}\wedge e_{\theta_2} - e_{\phi_1}\wedge e_{\phi_2}) \\
&- i\left[i(\alpha^2 - \beta^2)\sin\psi - \cos\psi\right]\mathcal{G}_3\sqrt{\mathcal{G}_2}\,e_\psi\wedge(e_{\phi_1}\wedge e_{\theta_2} + e_{\theta_1}\wedge e_{\phi_2}) \\
&- i\left[i(\alpha^2 - \beta^2)\cos\psi + \sin\psi\right]\mathcal{G}_3\sqrt{\mathcal{G}_2}\,e_\psi\wedge(e_{\theta_1}\wedge e_{\theta_2} - e_{\phi_1}\wedge e_{\phi_2}).
\end{aligned}$$
(3.36)

Plugging in the explicit values of $\alpha$, $\beta$, and $\mathcal{G}_i$ from (3.29) and (3.30), $\Omega$ is expressed as

$$
\begin{aligned}
\Omega &= \frac{i\mu^2 \mathbf{T}}{r^2} (e_{\theta_2} \wedge e_{\phi_2} - e_{\theta_1} \wedge e_{\phi_1}) \wedge \left( \frac{2i}{r\mathbf{S}} e_r + e_\psi \right) \\
&\quad - i\mathbf{T} (i\mathbf{S} \sin\psi - \cos\psi) \ e_\psi \wedge (e_{\phi_1} \wedge e_{\theta_2} + e_{\theta_1} \wedge e_{\phi_2}) \\
&\quad - i\mathbf{T} (i\mathbf{S} \cos\psi + \sin\psi) \ e_\psi \wedge (e_{\theta_1} \wedge e_{\theta_2} - e_{\phi_1} \wedge e_{\phi_2}) \\
&\quad + \frac{2i\mathbf{T}}{r\mathbf{S}} (\mathbf{S} \sin\psi + i\cos\psi) \ e_r \wedge (e_{\phi_1} \wedge e_{\theta_2} + e_{\theta_1} \wedge e_{\phi_2}) \\
&\quad + \frac{2i\mathbf{T}}{r\mathbf{S}} (\mathbf{S} \cos\psi - i\sin\psi) \ e_r \wedge (e_{\theta_1} \wedge e_{\theta_2} - e_{\phi_1} \wedge e_{\phi_2}) ,
\end{aligned}
\quad (3.37)
$$

where $\mathbf{S}$ and $\mathbf{T}$ are defined as

$$
\mathbf{S} \equiv \sqrt{1 - \frac{\mu^4}{r^4}}, \qquad \mathbf{T} \equiv \frac{\gamma}{8} \sqrt{\gamma + (r^2 \gamma' - \gamma)\left(1 - \frac{\mu^4}{r^4}\right)}.
\quad (3.38)
$$

We can now compute $d\Omega$, where we regularly use the identities (3.26):

$$
\begin{aligned}
d\Omega = &\left[ i\mu^2 \partial_r \left( \frac{\mathbf{T}}{r^2} \right) e_r \wedge (e_{\theta_2} \wedge e_{\phi_2} - e_{\theta_1} \wedge e_{\phi_1}) \wedge e_\psi \right. \\
&\left. + \frac{i\mu^2 \mathbf{T}}{r^2} (e_{\theta_2} \wedge e_{\phi_2} - e_{\theta_1} \wedge e_{\phi_1}) \wedge (-e_{\theta_1} \wedge e_{\phi_1} - e_{\theta_2} \wedge e_{\phi_2}) \right] \\
&\left[ +(\partial_r(\mathbf{TS})\sin\psi + i\partial_r(\mathbf{T})\cos\psi) e_r \wedge e_\psi \wedge (e_{\phi_1} \wedge e_{\theta_2} + e_{\theta_1} \wedge e_{\phi_2}) \right. \\
&\quad - i\mathbf{T}(i\mathbf{S}\cos\psi + \sin\psi) \ d\psi \wedge (\cos\theta_1 d\phi_1 + \cos\theta_2 d\phi_2) \wedge (e_{\phi_1} \wedge e_{\theta_2} + e_{\theta_1} \wedge e_{\phi_2}) \\
&\left. + i\mathbf{T}(i\mathbf{S}\sin\psi - \cos\psi) \ e_\psi \wedge (\cot\theta_1 e_{\theta_1} \wedge e_{\phi_1} \wedge e_{\theta_2} - \cot\theta_2 e_{\theta_1} \wedge e_{\theta_2} \wedge e_{\phi_2}) \right] \\
&\left[ -i\mathbf{T}(-i\mathbf{S}\sin\psi + \cos\psi) \ d\psi \wedge (\cos\theta_1 d\phi_1 + \cos\theta_2 d\phi_2) \wedge (e_{\theta_1} \wedge e_{\theta_2} - e_{\phi_1} \wedge e_{\phi_2}) \right. \\
&\quad + (\partial_r(\mathbf{TS})\cos\psi - i\partial_r(\mathbf{T})\sin\psi) e_r \wedge e_\psi \wedge (e_{\theta_1} \wedge e_{\theta_2} - e_{\phi_1} \wedge e_{\phi_2}) \\
&\left. + i\mathbf{T}(i\mathbf{S}\cos\psi + \sin\psi) \ e_\psi \wedge (-\cot\theta_1 e_{\theta_1} \wedge e_{\phi_1} \wedge e_{\phi_2} + \cot\theta_2 e_{\theta_1} \wedge e_{\theta_2} \wedge e_{\phi_2}) \right] \\
&\left[ + \frac{2i\mathbf{T}}{r\mathbf{S}} (\mathbf{S}\cos\psi - i\sin\psi) \ d\psi \wedge e_r \wedge (e_{\phi_1} \wedge e_{\theta_2} + e_{\theta_1} \wedge e_{\phi_2}) \right. \\
&\left. - \frac{2i\mathbf{T}}{r\mathbf{S}} (\mathbf{S}\sin\psi + i\cos\psi) \ e_r \wedge (\cot\theta_1 e_{\theta_1} \wedge e_{\phi_1} \wedge e_{\theta_2} - \cot\theta_2 e_{\theta_1} \wedge e_{\theta_2} \wedge e_{\phi_2}) \right]
\end{aligned}
$$

$$\left[+\frac{2i\mathbf{T}}{r\mathbf{S}}\left(-\mathbf{S}\sin\psi - i\cos\psi\right)d\psi \wedge\ e_r \wedge (e_{\theta_1} \wedge e_{\theta_2} - e_{\phi_1} \wedge e_{\phi_2})\right.$$
$$\left.-\frac{2i\mathbf{T}}{r\mathbf{S}}\left(\mathbf{S}\cos\psi - i\sin\psi\right)\ e_r \wedge \left(-\cot\theta_1 e_{\theta_1}\wedge e_{\phi_1}\wedge e_{\phi_2} + \cot\theta_2 e_{\phi_1}\wedge e_{\theta_2}\wedge e_{\phi_2}\right)\right].$$

Using the antisymmetric properties of the wedge product, collecting terms and defining the variables

$$\mathbf{E}_1 \equiv i\mu^2 \partial_r\left(\frac{\mathbf{T}}{r^2}\right), \qquad \mathbf{E}_2 \equiv \partial_r\left(\mathbf{ST}\right) - \frac{2\mathbf{T}}{r\mathbf{S}}, \qquad \mathbf{E}_3 \equiv \partial_r \mathbf{T} - \frac{2\mathbf{T}}{r}, \qquad (3.39)$$

$d\Omega$ simplifies to

$$\begin{aligned}
d\Omega &= \mathbf{E}_1\ e_r \wedge d\psi \wedge (e_{\theta_2}\wedge e_{\phi_2} - e_{\theta_1}\wedge e_{\phi_1}) & (3.40)\\
&+ (\mathbf{E}_2 \cos\psi - i\mathbf{E}_3 \sin\psi)\ e_r \wedge d\psi \wedge (e_{\theta_1}\wedge e_{\theta_2} - e_{\phi_1}\wedge e_{\phi_2})\\
&+ (\mathbf{E}_2 \sin\psi + i\mathbf{E}_3 \cos\psi)\ e_r \wedge d\psi \wedge (e_{\phi_1}\wedge e_{\theta_2} + e_{\theta_1}\wedge e_{\phi_2})\\
&+ (\mathbf{E}_2 \cos\psi - i\mathbf{E}_3 \sin\psi)\ e_r \wedge e_{\theta_1}\wedge e_{\theta_2} \wedge (\cot\ \theta_1\ e_{\phi_1} + \cot\ \theta_2\ e_{\phi_2})\\
&+ \left[\mathbf{E}_1\ \cot\ \theta_1 + \cot\ \theta_2\left(\mathbf{E}_2 \sin\psi + i\mathbf{E}_3 \cos\psi\right)\right]\ e_r \wedge e_{\phi_1}\wedge e_{\theta_2}\wedge e_{\phi_2}\\
&+ \left[\mathbf{E}_1\ \cot\ \theta_2 + \cot\ \theta_1\left(\mathbf{E}_2 \sin\psi + i\mathbf{E}_3 \cos\psi\right)\right]\ e_r \wedge e_{\phi_1}\wedge e_{\theta_1}\wedge e_{\phi_2}.
\end{aligned}$$

So in order for this space to be a complex manifold, we require $d\Omega$ to vanish. Staring at (3.40), we see that we need all the $\mathbf{E}_i$ coefficients to vanish. Since $\mathbf{E}_1 = 0$ implies $\partial_r\left(\frac{\mathbf{T}}{r^2}\right) = 0$, we have that $\mathbf{T}$ is proportional to $r^2$. Plugging this solution for $\mathbf{T}$ into the differential equation (3.33) implies the vanishing of the $\mathbf{E}_2$ and $\mathbf{E}_3$ coefficients as well, and so we have verified that the Kähler deformed conifold is complex.

Having found our formalism gives a consistent solution in the simpler Kähler case, we are ready to compute $d\Omega$ in the more general case (3.34):

$$d\Omega = e_r \wedge d\psi \wedge \left[\partial_\psi \mathbf{A}_{21}\ e_{\phi_1}\wedge e_{\theta_2} + \partial_\psi \mathbf{A}_{22}\ e_{\theta_1}\wedge e_{\phi_2} + \partial_\psi \mathbf{A}_{31}\ e_{\theta_1}\wedge e_{\theta_2} - \partial_\psi \mathbf{A}_{32}\ e_{\phi_1}\wedge e_{\phi_2}\right]$$

$$+ (\cos\theta_1 d\phi_1 + \cos\theta_2 d\phi_2) \wedge d\psi \wedge \left[\partial_\psi \mathbf{A}_{41} \; e_{\theta_2} \wedge e_{\phi_2} - \partial_\psi \mathbf{A}_{42} \; e_{\theta_1} \wedge e_{\phi_1}\right.$$

$$\left.+ \partial_\psi \mathbf{A}_{51} \; e_{\phi_1} \wedge e_{\theta_2} + \partial_\psi \mathbf{A}_{52} \; e_{\theta_1} \wedge e_{\phi_2} + \partial_\psi \mathbf{A}_{61} \; e_{\theta_1} \wedge e_{\theta_2}\right]$$

$$- e_\psi \wedge e_r \wedge \left[\partial_r \mathbf{A}_{41} \; e_{\theta_2} \wedge e_{\phi_2} - \partial_r \mathbf{A}_{42} \; e_{\theta_1} \wedge e_{\phi_1}\right.$$

$$\left.+ \partial_r \mathbf{A}_{51} \; e_{\phi_1} \wedge e_{\theta_2} + \partial_r \mathbf{A}_{52} \; e_{\theta_1} \wedge e_{\phi_2} + \partial_r \mathbf{A}_{61} \; e_{\theta_1} \wedge e_{\theta_2} + \partial_r \mathbf{A}_{61} e_{\theta_1} \wedge e_{\theta_2}\right]$$

$$\left[+ e_r \wedge (\mathbf{A}_{21} \; \cot\theta_1 e_{\theta_1} \wedge e_{\phi_1} \wedge e_{\theta_2} - \mathbf{A}_{22} \; e_{\theta_1} \wedge \cot\theta_2 e_{\theta_2} \wedge e_{\phi_2})\right.$$

$$+ e_r \wedge (-\mathbf{A}_{32} \; \cot\theta_1 e_{\theta_1} \wedge e_{\phi_1} \wedge e_{\phi_2} + \mathbf{A}_{32} \; e_{\phi_1} \wedge \cot\theta_2 e_{\theta_2} \wedge e_{\phi_2})$$

$$+ e_\psi \wedge (\mathbf{A}_{51} \; \cot\theta_1 e_{\theta_1} \wedge e_{\phi_1} \wedge e_{\theta_2} - \mathbf{A}_{52} \; e_{\theta_1} \wedge \cot\theta_2 e_{\theta_2} \wedge e_{\phi_2})$$

$$\left.+ e_\psi \wedge (-\mathbf{A}_{62} \; \cot\theta_1 e_{\theta_1} \wedge e_{\phi_1} \wedge e_{\phi_2} + \mathbf{A}_{62} \; e_{\phi_1} \wedge \cot\theta_2 e_{\theta_2} \wedge e_{\phi_2})\right]$$

$$- (e_{\theta_1} \wedge e_{\phi_1} + e_{\theta_2} \wedge e_{\phi_2}) \wedge \left[\mathbf{A}_{41} \; e_{\theta_2} \wedge e_{\phi_2} - \mathbf{A}_{42} \; e_{\theta_1} \wedge e_{\phi_1}\right.$$

$$\left.+ \mathbf{A}_{51} \; e_{\phi_1} \wedge e_{\theta_2} + \mathbf{A}_{52} \; e_{\theta_1} \wedge e_{\phi_2} + \mathbf{A}_{61} \; e_{\theta_1} \wedge e_{\theta_2} - \mathbf{A}_{62} \; e_{\phi_1} \wedge e_{\phi_2}\right].$$

Recalling that $\mathbf{A}_{nk} \equiv -i \; \mathbf{A}_{n-3,k}\sqrt{\frac{\mathcal{G}_2}{\mathcal{G}_1}}, n \geq 4$, we can define a new set of coefficients that simplify the above form once terms are collected:

$$\mathbf{D}_{1k} = \cot\;\theta_{k+s}\left[\mathbf{A}_{32} + i\;\partial_r\left(\mathbf{A}_{2k}\sqrt{\frac{\mathcal{G}_2}{\mathcal{G}_1}}\right)\right] + i\;\cot\;\theta_k\;\partial_r\left(\mathbf{A}_{1k}\sqrt{\frac{\mathcal{G}_2}{\mathcal{G}_1}}\right) \tag{3.41}$$

$$\mathbf{B}_{3k} = (-1)^k\left[\partial_\psi \mathbf{A}_{3k} + i\partial_r\left(\mathbf{A}_{3k}\sqrt{\frac{\mathcal{G}_2}{\mathcal{G}_1}}\right)\right], \quad \mathbf{C}_{1k} = \mathbf{A}_{2k} - i\partial_r\left(\mathbf{A}_{31}\sqrt{\frac{\mathcal{G}_2}{\mathcal{G}_1}}\right),$$

$$\mathbf{D}_{2k} = \sqrt{\frac{\mathcal{G}_2}{\mathcal{G}_1}}\left(\partial_\psi \mathbf{A}_{2k} - \mathbf{A}_{32}\right), \quad \mathbf{E} = \sqrt{\frac{\mathcal{G}_2}{\mathcal{G}_1}}\left[\mathbf{A}_{11} - \mathbf{A}_{12} + \cot\;\theta_1\;\cot\;\theta_2\;(\mathbf{A}_{22} - \mathbf{A}_{21})\right]$$

$$\mathbf{B}_{1k} = i(-1)^k\partial_r\left(\mathbf{A}_{1k}\sqrt{\frac{\mathcal{G}_2}{\mathcal{G}_1}}\right), \quad \mathbf{B}_{2k} = -\partial_\psi \mathbf{A}_{2k} - i\partial_r\left(\mathbf{A}_{2k}\sqrt{\frac{\mathcal{G}_2}{\mathcal{G}_1}}\right), \quad \mathbf{C}_{2k} = \frac{\mathbf{A}_{2k} + \partial_\psi \mathbf{A}_{31}}{\sqrt{\mathcal{G}_1/\mathcal{G}_2}},$$

where $s = (-1)^{k+1}$ and $k = 1, 2$. We then see that $d\Omega$ simplifies to

$$d\Omega = e_r \wedge e_{\theta_1} \wedge e_{\theta_2} \wedge (\mathbf{C}_{11} \; \cot\;\theta_1\; e_{\phi_1} + \mathbf{C}_{12}\; \cot\;\theta_2\; e_{\phi_2}) \tag{3.42}$$

$$
\begin{aligned}
&- \; i \, d\psi \wedge e_{\theta_1} \wedge e_{\theta_2} \wedge (\mathbf{C}_{21} \, \cot \, \theta_1 \, e_{\phi_1} + \mathbf{C}_{22} \, \cot \, \theta_2 \, e_{\phi_2}) \\
&+ \; e_r \wedge d\psi \wedge (\mathbf{B}_{31} \, e_{\theta_1} \wedge e_{\theta_2} + \mathbf{B}_{32} \, e_{\phi_1} \wedge e_{\phi_2}) + i \, \mathbf{E} \, e_{\theta_1} \wedge e_{\phi_1} \wedge e_{\theta_2} \wedge e_{\phi_2} \\
&+ \; e_r \wedge d\psi \wedge (\mathbf{B}_{11} \, e_{\theta_2} \wedge e_{\phi_2} + \mathbf{B}_{12} \, e_{\theta_1} \wedge e_{\phi_1} + \mathbf{B}_{21} \, e_{\phi_1} \wedge e_{\theta_2} + \mathbf{B}_{22} \, e_{\theta_1} \wedge e_{\phi_2}) \\
&+ \; e_{\phi_1} \wedge e_{\phi_2} \wedge [e_r \wedge (\mathbf{D}_{11} \, e_{\theta_2} + \mathbf{D}_{12} \, e_{\theta_1}) + i \, d\psi \wedge (\mathbf{D}_{21} \, \cot \, \theta_2 \, e_{\theta_2} + \mathbf{D}_{22} \, \cot \, \theta_1 \, e_{\theta_1})].
\end{aligned}
$$

Demanding that (3.42) vanishes then implies that each coefficient defined in (3.41) vanishes. Starting with $\mathbf{E} = 0$, we see that

$$\mathbf{A}_{22} = \mathbf{A}_{21}, \quad \mathbf{A}_{11} = \mathbf{A}_{12}. \tag{3.43}$$

$\mathbf{B}_{1k} = 0$ implies that the $\mathbf{A}_{1k}$ terms are proportional to $\sqrt{\frac{\mathcal{G}_1}{\mathcal{G}_2}}$:

$$\mathbf{A}_{11} = \mathbf{A}_{12} = \alpha \sqrt{\frac{\mathcal{G}_1}{\mathcal{G}_2}}, \tag{3.44}$$

where $\alpha$ is the constant of proportionality which is still undetermined. Taking into account equations (3.43) and (3.44), the vanishing of the $\mathbf{C}_{ik}$ and $\mathbf{D}_{ik}$ terms implies the four equations

$$
\begin{aligned}
\mathbf{A}_{21} - i \, \partial_r \left( \mathbf{A}_{31} \sqrt{\frac{\mathcal{G}_2}{\mathcal{G}_1}} \right) &= 0 \\
\mathbf{A}_{32} - \partial_\psi \mathbf{A}_{21} &= 0 \\
\mathbf{A}_{32} + i \, \partial_r \left( \mathbf{A}_{21} \sqrt{\frac{\mathcal{G}_2}{\mathcal{G}_1}} \right) &= 0 \\
\mathbf{A}_{21} + \partial_\psi \mathbf{A}_{31} &= 0.
\end{aligned}
\tag{3.45}
$$

Subtracting the first and fourth equations yields

$$\partial_\psi \mathbf{A}_{31} + i \partial_r \left( \mathbf{A}_{31} \sqrt{\frac{\mathcal{G}_2}{\mathcal{G}_1}} \right). \tag{3.46}$$

Subtracting the second and third equations yields

$$\partial_\psi \mathbf{A}_{21} + i\partial_r \left( \mathbf{A}_{21} \sqrt{\frac{\mathcal{G}_2}{\mathcal{G}_1}} \right). \tag{3.47}$$

Taking $\partial_\psi$ of the third equation we have

$$\partial_\psi \mathbf{A}_{32} + i\partial_r \left( \partial_\psi \mathbf{A}_{21} \sqrt{\frac{\mathcal{G}_2}{\mathcal{G}_1}} \right) = \partial_\psi \mathbf{A}_{32} + i\partial_r \left( \mathbf{A}_{32} \sqrt{\frac{\mathcal{G}_2}{\mathcal{G}_1}} \right), \tag{3.48}$$

where we used the second equation to replace $\partial_\psi \mathbf{A}_{21}$ with $\mathbf{A}_{32}$. These three equations are exactly the three equations implied by the vanishing of the $\mathbf{B}_{2k}$ and $\mathbf{B}_{3k}$ coefficients. So the set of equations are consistent, and so metrics with coefficients satisfying the constraints (3.43), (3.44), (3.46), (3.47), and (3.48) will describe complex manifolds.

We can test these conditions in the case of the warped-resolved conifold. Plugging in the values from (3.15) into the definitions of the $\mathbf{A}_{ij}$ from (3.35) yields

$$\mathbf{A}_{11} = \mathbf{A}_{12} = 0, \quad \mathbf{A}_{31} = \mathbf{A}_{32} = -i\,\mathbf{A}_{21} = -i\,\mathbf{A}_{22} = \sqrt{\mathcal{G}_1 \mathcal{G}_3 \mathcal{G}_4}\,(i\cos\psi + \sin\psi)\,.$$

We immediately see that (3.43) and (3.44) are satisfied. Plugging any of the $\mathbf{A}_{31}, \mathbf{A}_{32}, -i\,\mathbf{A}_{21}$ into their respective differential equations (3.46),(3.47), and (3.48) yields the differential equation

$$\sqrt{\mathcal{G}_1 \mathcal{G}_3 \mathcal{G}_4}(-i\sin\psi + \cos\psi) + (\sqrt{\mathcal{G}_2 \mathcal{G}_3 \mathcal{G}_4})_r(-\cos\psi + i\sin\psi) = 0$$
$$\implies \frac{\mathcal{G}_{3r}}{\mathcal{G}_3} + \frac{\mathcal{G}_{4r}}{\mathcal{G}_4} + \frac{\mathcal{G}_{2r} - 2\sqrt{\mathcal{G}_1 \mathcal{G}_2}}{\mathcal{G}_2} = 0, \tag{3.49}$$

which exactly matches the equation for the $\mathcal{G}_i$ derived in [4].

## 3.3 Computing the complex 3-form $\mathbf{G}_3$

The goal of this section will be to calculate the complex three form $\mathbf{G}_3 = \mathbf{F}_3 - ie^{-i\phi}\mathbf{H}_3$ in terms of complex vielbeins of the post-dual manifold. This calculation involves taking the hodge star operator with respect to $ds_6^2$, so it is important to rewrite our expression for $d(e^{2\phi}J)$ from (3.2) in terms of the vielbeins. In order to do so we will need an expression for the basis vectors $e_r, e_\psi, e_{\theta_i}, e_{\phi_i}$ in terms of the vielbeins $e_1, \ldots, e_6$.

We begin by writing the basis vectors (3.19) in terms of the left invariant Maurer-Cartan forms $\sigma_i, \Sigma_i$, $i = 1, 2$. To do this we first rewrite (3.20) as a set of matrix equations:

$$\begin{pmatrix} \sigma_1 \\ \sigma_2 \end{pmatrix} = \begin{pmatrix} \cos\frac{\psi}{2} & \sin\frac{\psi}{2} \\ -\sin\frac{\psi}{2} & \cos\frac{\psi}{2} \end{pmatrix} \begin{pmatrix} e_{\theta_1} \\ e_{\phi_1} \end{pmatrix}, \quad \begin{pmatrix} \Sigma_1 \\ \Sigma_2 \end{pmatrix} = \begin{pmatrix} \cos\frac{\psi}{2} & \sin\frac{\psi}{2} \\ -\sin\frac{\psi}{2} & \cos\frac{\psi}{2} \end{pmatrix} \begin{pmatrix} e_{\theta_2} \\ e_{\phi_2} \end{pmatrix}.$$

Inverting these equations yields

$$\begin{pmatrix} e_{\theta_1} \\ e_{\phi_1} \end{pmatrix} = \begin{pmatrix} \cos\frac{\psi}{2} & -\sin\frac{\psi}{2} \\ \sin\frac{\psi}{2} & \cos\frac{\psi}{2} \end{pmatrix} \begin{pmatrix} \sigma_1 \\ \sigma_2 \end{pmatrix}, \quad \begin{pmatrix} e_{\theta_2} \\ e_{\phi_2} \end{pmatrix} = \begin{pmatrix} \cos\frac{\psi}{2} & -\sin\frac{\psi}{2} \\ \sin\frac{\psi}{2} & \cos\frac{\psi}{2} \end{pmatrix} \begin{pmatrix} \Sigma_1 \\ \Sigma_2 \end{pmatrix}.$$

(3.50)

Recall from (3.8) that

$$e_1 = \sqrt{\mathcal{G}_1} e_r, \quad e_2 = \sqrt{\mathcal{G}_2} e_\psi, \quad e_3 = \sqrt{\mathcal{G}_3}\left(\alpha_1 \sigma_1 + \beta_3 \Sigma_1\right), \quad e_4 = \sqrt{\mathcal{G}_4}\left(\alpha_2 \sigma_2 - \beta_4 \Sigma_2\right)$$
$$e_5 = \sqrt{\mathcal{G}_3}\left(\beta_1 \sigma_1 + \alpha_3 \Sigma_1\right), \quad e_6 = \sqrt{\mathcal{G}_4}\left(-\beta_2 \sigma_2 + \alpha_4 \Sigma_2\right).$$

Staring at these equations, we see we can isolate each $\sigma_i$ and $\Sigma_i$ by taking the following clever linear combinations of pairs of $e_i$:

$$\sigma_1 = \frac{\alpha_3 e_3 - \beta_3 e_5}{(\alpha_1 \alpha_3 - \beta_1 \beta_3)\sqrt{\mathcal{G}_3}}, \quad \sigma_2 = \frac{\alpha_4 e_4 + \beta_4 e_6}{(\alpha_2 \alpha_4 - \beta_2 \beta_4)\sqrt{\mathcal{G}_4}}$$

$$\Sigma_1 = \frac{\alpha_1 e_5 - \beta_1 e_3}{(\alpha_1 \alpha_3 - \beta_1 \beta_3) \sqrt{\mathcal{G}_3}}, \qquad \Sigma_2 = \frac{\alpha_2 e_6 + \beta_2 e_4}{(\alpha_2 \alpha_4 - \beta_2 \beta_4) \sqrt{\mathcal{G}_4}}. \tag{3.51}$$

We can now write the basis vectors in terms of the vielbeins by combining (3.51) and (3.50):

$$\begin{aligned}
e_{\theta_1} &= \cos\frac{\psi}{2} \frac{\alpha_3 e_3 - \beta_3 e_5}{(\alpha_1 \alpha_3 - \beta_1 \beta_3) \sqrt{\mathcal{G}_3}} - \sin\frac{\psi}{2} \frac{\alpha_4 e_4 + \beta_4 e_6}{(\alpha_2 \alpha_4 - \beta_2 \beta_4) \sqrt{\mathcal{G}_4}} \\
e_{\phi_1} &= \sin\frac{\psi}{2} \frac{\alpha_3 e_3 - \beta_3 e_5}{(\alpha_1 \alpha_3 - \beta_1 \beta_3) \sqrt{\mathcal{G}_3}} + \cos\frac{\psi}{2} \frac{\alpha_4 e_4 + \beta_4 e_6}{(\alpha_2 \alpha_4 - \beta_2 \beta_4) \sqrt{\mathcal{G}_4}} \\
e_{\theta_2} &= \cos\frac{\psi}{2} \frac{\alpha_1 e_5 - \beta_1 e_3}{(\alpha_1 \alpha_3 - \beta_1 \beta_3) \sqrt{\mathcal{G}_3}} - \sin\frac{\psi}{2} \frac{\alpha_2 e_6 + \beta_2 e_4}{(\alpha_2 \alpha_4 - \beta_2 \beta_4) \sqrt{\mathcal{G}_4}} \\
e_{\phi_2} &= \sin\frac{\psi}{2} \frac{\alpha_1 e_5 - \beta_1 e_3}{(\alpha_1 \alpha_3 - \beta_1 \beta_3) \sqrt{\mathcal{G}_3}} + \cos\frac{\psi}{2} \frac{\alpha_2 e_6 + \beta_2 e_4}{(\alpha_2 \alpha_4 - \beta_2 \beta_4) \sqrt{\mathcal{G}_4}}
\end{aligned} \tag{3.52}$$

We can now use (3.52) to write $d(e^{2\phi} J)$ solely in terms of vielbeins. However we first need to compute $d(e^{2\phi} J)$ without making the $\mathbf{A} = \mathbf{B}$ simplification in order to proceed with the calculation of $\mathbf{G}_3$ in full generality. We define the following coefficients to simplify the expression of $e^{2\phi} J$ before taking the exterior derivative:

$$\mathbf{K} \equiv -\frac{1}{2} \sqrt{\mathcal{G}_3 \mathcal{G}_4} \, (\mathbf{A} - \mathbf{B}) \sin\psi \tag{3.53}$$

$$\mathcal{F}_1 \equiv \left( \mathbf{A} \cos^2\frac{\psi}{2} + \mathbf{B} \sin^2\frac{\psi}{2} \right) \sqrt{\mathcal{G}_3 \mathcal{G}_4}, \qquad \mathcal{F}_2 \equiv \left( \mathbf{B} \cos^2\frac{\psi}{2} + \mathbf{A} \sin^2\frac{\psi}{2} \right) \sqrt{\mathcal{G}_3 \mathcal{G}_4}.$$

$e^{2\phi} J$ now reads

$$\begin{aligned}
e^{2\phi} J &= \sqrt{\mathcal{G}_1 \mathcal{G}_2} dr \wedge e_\psi \\
&+ \mathbf{K} \left( e_{\phi_1} \wedge e_{\phi_2} - e_{\theta_1} \wedge e_{\theta_2} \right) \\
&- \mathbf{C} e_{\theta_1} \wedge e_{\phi_1} - \mathbf{D} e_{\theta_2} \wedge e_{\phi_2} \\
&- \mathcal{F}_1 e_{\theta_1} \wedge e_{\phi_2} - \mathcal{F}_2 e_{\theta_2} \wedge e_{\phi_1},
\end{aligned} \tag{3.54}$$

where **C** and **D** are defined as in (3.24). Again, making use of the identities (3.26), we have

$$\begin{aligned}
d(e^{2\phi}J) &= \sqrt{\mathcal{G}_1\mathcal{G}_2}e_r \wedge (e_{\theta_1} \wedge e_{\phi_1} + e_{\theta_2} \wedge e_{\phi_2}) \\
&+ (\mathbf{K}_r e_r + \mathbf{K}_\psi d\psi) \wedge (e_{\phi_1} \wedge e_{\phi_2} - e_{\theta_1} \wedge e_{\theta_2}) \\
&+ \mathbf{K}(\cot\theta_1 e_{\theta_1} \wedge e_{\phi_1} \wedge e_{\phi_2} - \cot\theta_2 e_{\phi_1} \wedge e_{\theta_2} \wedge e_{\phi_2}) \\
&- \mathbf{C}_r e_r \wedge e_{\theta_1} \wedge e_{\phi_1} - \mathbf{D}_r e_r \wedge e_{\theta_2} \wedge e_{\phi_2} \\
&- (\mathcal{F}_{1,r} e_r + \mathcal{F}_{1,\psi} d\psi) \wedge e_{\theta_1} \wedge e_{\phi_2} + \mathcal{F}_1 \cot\theta_2 e_{\theta_1} \wedge e_{\theta_2} \wedge e_{\phi_2} \\
&- (\mathcal{F}_{2,r} e_r + \mathcal{F}_{2,\psi} d\psi) \wedge e_{\theta_2} \wedge e_{\phi_1} + \mathcal{F}_2 \cot\theta_1 e_{\theta_2} \wedge e_{\theta_1} \wedge e_{\phi_1}.
\end{aligned}$$

Defining the coefficients

$$\mathbf{K}_3 \equiv (\mathbf{K}_\psi - \mathcal{F}_2)\cot\theta_1, \quad \mathbf{K}_4 \equiv (\mathbf{K}_\psi + \mathcal{F}_1)\cot\theta_2$$

$$\mathbf{K}_5 \equiv (\mathbf{K} - \mathcal{F}_{1\psi})\cot\theta_1, \quad \mathbf{K}_6 \equiv (\mathbf{K} + \mathcal{F}_{2\psi})\cot\theta_2, \tag{3.55}$$

the closure of $J$ simplifies to

$$\begin{aligned}
d\left(e^{2\phi}J\right) &= (e_{\phi_1} \wedge e_{\phi_2} - e_{\theta_1} \wedge e_{\theta_2}) \wedge (\mathbf{K}_r e_r + \mathbf{K}_\psi e_\psi) \\
&+ e_{\theta_1} \wedge e_{\theta_2} \wedge (\mathbf{K}_3 e_{\phi_1} + \mathbf{K}_4 e_{\phi_2}) + e_{\phi_1} \wedge e_{\phi_2} \wedge (\mathbf{K}_5 e_{\theta_1} + \mathbf{K}_6 e_{\theta_2}) \\
&+ e_r \wedge \left[e_{\theta_1} \wedge e_{\phi_1}\left(\sqrt{\mathcal{G}_1\mathcal{G}_2} - \mathbf{C}_r\right) + e_{\theta_2} \wedge e_{\phi_2}\left(\sqrt{\mathcal{G}_1\mathcal{G}_2} - \mathbf{D}_r\right)\right] \\
&- \mathcal{F}_{1r} e_r \wedge \left(e_{\theta_1} \wedge e_{\phi_2} + \frac{\mathcal{F}_{2r}}{\mathcal{F}_{1r}} e_{\theta_2} \wedge e_{\phi_1}\right) - \mathcal{F}_{1\psi} e_\psi \wedge \left(e_{\theta_1} \wedge e_{\phi_2} + \frac{\mathcal{F}_{2\psi}}{\mathcal{F}_{1\psi}} e_{\theta_2} \wedge e_{\phi_1}\right).
\end{aligned} \tag{3.56}$$

We can anticipate a large number of possible coefficients of $d(e^{2\phi}J)$ when we expand it in the basis of vielbeins $e_i$ by comparing (3.56) and (3.52). We will explicitly show the procedure for the first line of (3.56). For the $\mathbf{K}_r$ term, expanding the basis vectors with respect to the vielbeins using (3.52) yields

$$\mathbf{K}_r \left(e_{\phi_1} \wedge e_{\phi_2} - e_{\theta_1} \wedge e_{\theta_2}\right) \wedge e_r = \frac{(\alpha_4\beta_1 + \alpha_3\beta_2)\mathbf{K}_r \sin\psi}{\sqrt{\mathcal{G}_1\mathcal{G}_3\mathcal{G}_4}\,(\alpha_1\alpha_3 - \beta_1\beta_3)(\alpha_2\alpha_4 - \beta_2\beta_4)} e_1 \wedge e_3 \wedge e_4$$
$$- \frac{\mathbf{K}_r \cos\psi}{\sqrt{\mathcal{G}_1}\mathcal{G}_3(\alpha_1\alpha_3 - \beta_1\beta_3)} e_1 \wedge e_3 \wedge e_5 + \frac{(\alpha_2\alpha_3 + \beta_1\beta_4)\mathbf{K}_r \sin\psi}{\sqrt{\mathcal{G}_1\mathcal{G}_3\mathcal{G}_4}(\alpha_1\alpha_3 - \beta_1\beta_3)(\alpha_2\alpha_4 - \beta_2\beta_4)} e_1 \wedge e_3 \wedge e_6$$
$$\frac{(\alpha_1\alpha_4 + \beta_2\beta_3)\mathbf{K}_r \sin\psi}{\sqrt{\mathcal{G}_1\mathcal{G}_3\mathcal{G}_4}(\alpha_1\alpha_3 - \beta_1\beta_3)(\alpha_2\alpha_4 - \beta_2\beta_4)} e_1 \wedge e_4 \wedge e_5 + \frac{\mathbf{K}_r \cos\psi}{\sqrt{\mathcal{G}_1}\mathcal{G}_4(\alpha_2\alpha_4 - \beta_2\beta_4)} e_1 \wedge e_4 \wedge e_6$$
$$- \frac{(\alpha_2\beta_3 + \alpha_1\beta_4)\mathbf{K}_r \sin\psi}{\sqrt{\mathcal{G}_1\mathcal{G}_3\mathcal{G}_4}(\alpha_1\alpha_3 - \beta_1\beta_3)(\alpha_2\alpha_4 - \beta_2\beta_4)} e_1 \wedge e_5 \wedge e_6,$$

where we used $e_r = e_1/\sqrt{\mathcal{G}_1}$. We can simplify this expression by making the following definitions:

$$\mathbf{H}_{3c}^{(0)} \equiv \frac{\mathbf{K}_r \sin\psi}{\sqrt{\mathcal{G}_1\mathcal{G}_3\mathcal{G}_4}(\alpha_1\alpha_3 - \beta_1\beta_3)(\alpha_2\alpha_4 - \beta_2\beta_4)}$$
$$\mathbf{H}_{3a}^{(0)} \equiv -\frac{\mathbf{K}_r \cos\psi}{(\alpha_1\alpha_3 - \beta_1\beta_3)\mathcal{G}_3\sqrt{\mathcal{G}_1}}, \quad \mathbf{H}_{3b}^{(0)} \equiv \frac{\mathbf{K}_r \cos\psi}{(\alpha_2\alpha_4 - \beta_2\beta_4)\mathcal{G}_4\sqrt{\mathcal{G}_1}}. \quad (3.57)$$

This now simplifies the $\mathbf{K}_r$ term to

$$\mathbf{K}_r \left(e_{\phi_1} \wedge e_{\phi_2} - e_{\theta_1} \wedge e_{\theta_2}\right) \wedge e_r$$
$$= (\alpha_4\beta_1 + \alpha_3\beta_2)\mathbf{H}_{3c}^{(0)} e_1 \wedge e_3 \wedge e_4 + \mathbf{H}_{3a}^{(0)} e_1 \wedge e_3 \wedge e_5 + (\alpha_2\alpha_3 + \beta_1\beta_4)\mathbf{H}_{3c}^{(0)} e_1 \wedge e_3 \wedge e_6$$
$$+ (\alpha_1\alpha_4 + \beta_2\beta_3)\mathbf{H}_{3c}^{(0)} e_1 \wedge e_4 \wedge e_5 + \mathbf{H}_{3b}^{(0)} e_1 \wedge e_4 \wedge e_6 - (\alpha_2\beta_3 + \alpha_1\beta_4)\mathbf{H}_{3c}^{(0)} e_1 \wedge e_5 \wedge e_6.$$
$$(3.58)$$

All that changes in the $\mathbf{K}_\psi$ term is that now we have $\mathbf{K}_\psi$ in the numerator instead of $\mathbf{K}_r$, and $\sqrt{\mathcal{G}_2}$ in the denominator instead of $\sqrt{\mathcal{G}_1}$ (since $e_\psi = \frac{e_2}{\sqrt{\mathcal{G}_2}}$). This implies that we will have a new set of coefficients $\mathbf{H}_{3n}^{'(0)}$ defined by

$$\mathbf{H}_{3n}^{'(0)} = \left(\frac{\mathbf{K}_\psi}{\mathbf{K}_r}\sqrt{\frac{\mathcal{G}_1}{\mathcal{G}_2}}\right)\mathbf{H}_{3n}^{(0)}, \quad n = a, b, c. \quad (3.59)$$

This demonstrates the procedure of defining the new coefficients that will appear in the final expression for $d(e^{2\phi}J)$. When considering the three other lines in (3.56), the following other coefficients appear:

$$\mathbf{H}_1^{(0)} \equiv \frac{\sqrt{\mathcal{G}_1\mathcal{G}_2} - \mathbf{C}_r}{\sqrt{\mathcal{G}_1\mathcal{G}_3\mathcal{G}_4}\,(\alpha_1\alpha_3 - \beta_1\beta_3)(\alpha_2\alpha_4 - \beta_2\beta_4)}, \qquad \mathbf{H}_2^{(0)} \equiv \left(\frac{\sqrt{\mathcal{G}_1\mathcal{G}_2} - \mathbf{D}_r}{\sqrt{\mathcal{G}_1\mathcal{G}_2} - \mathbf{C}_r}\right)\mathbf{H}_1^{(0)}$$

$$\mathbf{H}_{3c}^{(1)} \equiv \frac{\mathcal{F}_{1r}\sin^2\frac{\psi}{2} + \mathcal{F}_{2r}\cos^2\frac{\psi}{2}}{\sqrt{\mathcal{G}_1\mathcal{G}_3\mathcal{G}_4}\,(\alpha_1\alpha_3 - \beta_1\beta_3)(\alpha_2\alpha_4 - \beta_2\beta_4)}$$

$$\mathbf{H}_{3a}^{(1)} \equiv \frac{(\mathcal{F}_{2r} - \mathcal{F}_{1r})\sin\psi}{2(\alpha_1\alpha_3 - \beta_1\beta_3)\mathcal{G}_3\sqrt{\mathcal{G}_1}}, \qquad \mathbf{H}_{3b}^{(1)} \equiv -\frac{(\mathcal{F}_{2r} - \mathcal{F}_{1r})\sin\psi}{2(\alpha_2\alpha_4 - \beta_2\beta_4)\mathcal{G}_4\sqrt{\mathcal{G}_1}}$$

$$\mathbf{H}_{3c}^{\prime(1)} \equiv \frac{\mathcal{F}_{1\psi}\sin^2\frac{\psi}{2} + \mathcal{F}_{2\psi}\cos^2\frac{\psi}{2}}{\sqrt{\mathcal{G}_2\mathcal{G}_3\mathcal{G}_4}\,(\alpha_1\alpha_3 - \beta_1\beta_3)(\alpha_2\alpha_4 - \beta_2\beta_4)}$$

$$\mathbf{H}_{3k}^{\prime(1)} = \left(\frac{\mathcal{F}_{2\psi} - \mathcal{F}_{1\psi}}{\mathcal{F}_{2r} - \mathcal{F}_{1r}}\sqrt{\frac{\mathcal{G}_1}{\mathcal{G}_2}}\right)\mathbf{H}_{3k}^{(1)}, \quad k = a,b$$

$$b_o \equiv \frac{\mathcal{F}_{1r}\cos^2\frac{\psi}{2} + \mathcal{F}_{2r}\sin^2\frac{\psi}{2}}{\mathcal{F}_{1r}\sin^2\frac{\psi}{2} + \mathcal{F}_{2r}\cos^2\frac{\psi}{2}}, \qquad c_o \equiv \frac{\mathcal{F}_{1\psi}\cos^2\frac{\psi}{2} + \mathcal{F}_{2\psi}\sin^2\frac{\psi}{2}}{\mathcal{F}_{1\psi}\sin^2\frac{\psi}{2} + \mathcal{F}_{2\psi}\cos^2\frac{\psi}{2}}$$

$$\mathbf{H}_{4a}^{(0)} \equiv \frac{\mathbf{K}_3\sin\frac{\psi}{2}}{(\alpha_1\alpha_3 - \beta_1\beta_3)(\alpha_2\alpha_4 - \beta_2\beta_4)\mathcal{G}_4\sqrt{\mathcal{G}_3}}$$

$$\mathbf{H}_{4b}^{(0)} \equiv \frac{\mathbf{K}_3\cos\frac{\psi}{2}}{(\alpha_1\alpha_3 - \beta_1\beta_3)(\alpha_2\alpha_4 - \beta_2\beta_4)\mathcal{G}_3\sqrt{\mathcal{G}_4}}$$

$$\mathbf{H}_{4c}^{(0)} \equiv -\mathbf{H}_{4b}^{(0)}\left(\frac{\mathbf{K}_4}{\mathbf{K}_3}\right), \qquad \mathbf{H}_{4d}^{(0)} \equiv -\mathbf{H}_{4a}^{(0)}\left(\frac{\mathbf{K}_4}{\mathbf{K}_3}\right)$$

$$\mathbf{H}_{4a}^{\prime(0)} \equiv \mathbf{H}_{4a}^{(0)}\cot\frac{\psi}{2}\left(\frac{\mathbf{K}_5}{\mathbf{K}_3}\right), \qquad \mathbf{H}_{4b}^{\prime(0)} \equiv -\mathbf{H}_{4b}^{(0)}\tan\frac{\psi}{2}\left(\frac{\mathbf{K}_5}{\mathbf{K}_3}\right)$$

$$\mathbf{H}_{4c}^{\prime(0)} \equiv \mathbf{H}_{4b}^{(0)}\tan\frac{\psi}{2}\left(\frac{\mathbf{K}_6}{\mathbf{K}_3}\right), \qquad \mathbf{H}_{4d}^{\prime(0)} \equiv -\mathbf{H}_{4a}^{(0)}\cot\frac{\psi}{2}\left(\frac{\mathbf{K}_6}{\mathbf{K}_3}\right). \tag{3.60}$$

However, as can be seen, for example, in (3.58), the general coefficient in front of a given basis vector $e_i \wedge e_j \wedge e_k$ will be a linear combination of $\mathbf{H}$ coefficients. The final expression for $d(e^{2\phi}J)$ has the following 16 distinct linear combinations of the $H$ coefficients in front of the basis vectors after the whole expression is computed:

$$\mathbf{M}_1 \equiv \mathbf{H}_1^{(0)}\alpha_3\alpha_4 - \mathbf{H}_2^{(0)}\beta_1\beta_2 + \mathbf{H}_{3c}^{(0)}(\alpha_3\beta_2 + \alpha_4\beta_1) + \mathbf{H}_{3c}^{(1)}(\alpha_4\beta_1 - b_o\alpha_3\beta_2)$$

$$\mathbf{M}_2 \equiv \mathbf{H}_1^{(0)}\alpha_3\beta_4 - \mathbf{H}_2^{(0)}\beta_1\alpha_2 + \mathbf{H}_{3c}^{(0)}(\alpha_2\alpha_3 + \beta_1\beta_4) + \mathbf{H}_{3c}^{(1)}(\beta_1\beta_4 - b_o\alpha_2\alpha_3)$$

$$\mathbf{M}_3 \equiv \mathbf{H}_1^{(0)}\alpha_4\beta_3 - \mathbf{H}_2^{(0)}\beta_2\alpha_1 + \mathbf{H}_{3c}^{(0)}(\alpha_1\alpha_4 + \beta_2\beta_3) + \mathbf{H}_{3c}^{(1)}(\alpha_1\alpha_4 - b_o\beta_2\beta_3)$$

$$\mathbf{M}_4 \equiv -\mathbf{H}_1^{(0)}\beta_3\beta_4 + \mathbf{H}_2^{(0)}\alpha_1\alpha_2 - \mathbf{H}_{3c}^{(0)}(\alpha_1\beta_4 + \alpha_2\beta_3) - \mathbf{H}_{3c}^{(1)}(\alpha_1\beta_4 - b_o\beta_3\alpha_2)$$

$$\mathbf{Q}_1 \equiv \mathbf{H}_{3a}^{(0)} + \mathbf{H}_{3a}^{(1)}, \qquad \mathbf{Q}_2 \equiv \mathbf{H}_{3b}^{(0)} + \mathbf{H}_{3b}^{(1)}$$

$$\mathbf{Q}_3 \equiv \mathbf{H}_{3a}^{'(0)} + \mathbf{H}_{3a}^{'(1)}, \qquad \mathbf{Q}_4 \equiv \mathbf{H}_{3b}^{'(0)} + \mathbf{H}_{3b}^{'(1)}$$

$$\mathbf{P}_1 \equiv \mathbf{H}_{3c}^{'(0)}(\alpha_1\alpha_4 + \beta_2\beta_3) + \mathbf{H}_{3c}^{'(1)}(\alpha_1\alpha_4 - c_o\beta_2\beta_3)$$

$$\mathbf{P}_2 \equiv \mathbf{H}_{3c}^{'(0)}(\alpha_2\alpha_3 + \beta_1\beta_4) + \mathbf{H}_{3c}^{'(1)}(\beta_1\beta_4 - c_o\alpha_2\alpha_3)$$

$$\mathbf{P}_3 \equiv \mathbf{H}_{3c}^{'(0)}(\alpha_3\beta_2 + \alpha_4\beta_1) + \mathbf{H}_{3c}^{'(1)}(\beta_1\alpha_4 - c_o\alpha_3\beta_2)$$

$$\mathbf{P}_4 \equiv -\mathbf{H}_{3c}^{'(0)}(\alpha_1\beta_4 + \alpha_2\beta_3) - \mathbf{H}_{3c}^{'(1)}(\alpha_1\beta_4 - c_o\alpha_2\beta_3)$$

$$\mathbf{N}_1 \equiv \alpha_3\left(\mathbf{H}_{4a}^{(0)} + \mathbf{H}_{4a}^{'(0)}\right) + \beta_1\left(\mathbf{H}_{4d}^{(0)} + \mathbf{H}_{4d}^{'(0)}\right)$$

$$\mathbf{N}_2 \equiv \beta_3\left(\mathbf{H}_{4a}^{(0)} + \mathbf{H}_{4a}^{'(0)}\right) + \alpha_1\left(\mathbf{H}_{4d}^{(0)} + \mathbf{H}_{4d}^{'(0)}\right)$$

$$\mathbf{N}_3 \equiv \beta_2\left(\mathbf{H}_{4c}^{(0)} + \mathbf{H}_{4c}^{'(0)}\right) - \alpha_4\left(\mathbf{H}_{4b}^{(0)} + \mathbf{H}_{4b}^{'(0)}\right)$$

$$\mathbf{N}_4 \equiv \beta_4\left(\mathbf{H}_{4b}^{(0)} + \mathbf{H}_{4b}^{'(0)}\right) - \alpha_2\left(\mathbf{H}_{4c}^{(0)} + \mathbf{H}_{4c}^{'(0)}\right). \tag{3.61}$$

The expression for $d(e^{2\phi}J)$ now reads

$$d(e^{2\phi}J) = \mathbf{P}_4 e_2 \wedge e_5 \wedge e_6 + \mathbf{P}_1 e_2 \wedge e_4 \wedge e_5 + \mathbf{P}_2 e_2 \wedge e_3 \wedge e_6$$

$$+ \mathbf{Q}_2 e_1 \wedge e_4 \wedge e_6 + \mathbf{Q}_1 e_1 \wedge e_3 \wedge e_5 + \mathbf{M}_2 e_1 \wedge e_3 \wedge e_6$$

$$+ \mathbf{M}_3 e_1 \wedge e_4 \wedge e_5 + \mathbf{M}_4 e_1 \wedge e_5 \wedge e_6 + \mathbf{M}_1 e_1 \wedge e_3 \wedge e_4$$

$$+ \mathbf{Q}_4 e_2 \wedge e_4 \wedge e_6 + \mathbf{Q}_3 e_2 \wedge e_3 \wedge e_5 + \mathbf{N}_1 e_3 \wedge e_4 \wedge e_6$$

$$+ \mathbf{N}_2 e_4 \wedge e_5 \wedge e_6 + \mathbf{N}_3 e_3 \wedge e_4 \wedge e_5 + \mathbf{N}_4 e_3 \wedge e_5 \wedge e_6. \tag{3.62}$$

To compute the Hodge dual of the fundamental $(1,1)$ form, $*_6 d(e^{2\phi}J)$, we need a simple formula for that operation. Since we rewrote $d(e^{2\phi}J)$ in terms of vielbeins, this drastically simplifies the amount of computing needed. The general formula for the Hodge dual of a 3-form is given as follows:

$$*_6 \, \omega^3 = \sum\nolimits_{j_1<j_2<j_3}^{k_1<k_2<k_3} \sqrt{|\det g|} \omega^{k_1 k_2 k_3} \epsilon_{k_1 k_2 k_3 j_1 j_2 j_3} e_{j_1} \wedge e_{j_2} \wedge e_{j_3}. \tag{3.63}$$

Since our 3-form $d(e^{2\phi} J)$ is now written in a basis of vielbeins, the metric is given in components by $g_{ij} = \delta_{ij}$. This implies that $\det g = 1$ and $\omega^{k_1 k_2 k_3} = \omega_{k_1 k_2 k_3}$. So we can simplify our formula for the Hodge-dual of a three-form to

$$*_6 \, \omega^3 = \sum\nolimits_{j_1<j_2<j_3}^{k_1<k_2<k_3} \omega_{k_1 k_2 k_3} \epsilon_{k_1 k_2 k_3 j_1 j_2 j_3} e_{j_1} \wedge e_{j_2} \wedge e_{j_3}. \tag{3.64}$$

We can easily use this formula since our expression for $d(e^{2\phi} J)$ is written in a basis with increasing indices, as in the formula (3.64). For example, to find $[*_6 d(e^{2\phi} J)]_{135}$, we simply paste the $e_2 \wedge e_4 \wedge e_6$ component of $d(e^{2\phi} J)$ and we are right up to a sign. To determine the sign you simply compute $\epsilon_{135246}$, which happens to be $-1$. Now using our rule for the hodge star we derived in (3.64), we have

$$\begin{aligned}
*_6 d(e^{2\phi} J) = &\mathbf{M}_1 e_2 \wedge e_5 \wedge e_6 + \mathbf{M}_2 e_2 \wedge e_4 \wedge e_5 + \mathbf{M}_3 e_2 \wedge e_3 \wedge e_6 \\
&+ \mathbf{Q}_3 e_1 \wedge e_4 \wedge e_6 + \mathbf{Q}_4 e_1 \wedge e_3 \wedge e_5 - \mathbf{P}_1 e_1 \wedge e_3 \wedge e_6 \\
&- \mathbf{P}_2 e_1 \wedge e_4 \wedge e_5 - \mathbf{P}_3 e_1 \wedge e_5 \wedge e_6 - \mathbf{P}_4 e_1 \wedge e_3 \wedge e_4 \\
&- \mathbf{Q}_1 e_2 \wedge e_4 \wedge e_6 - \mathbf{Q}_2 e_2 \wedge e_3 \wedge e_5 - \mathbf{N}_1 e_1 \wedge e_2 \wedge e_5 \\
&- \mathbf{N}_2 e_1 \wedge e_2 \wedge e_3 + \mathbf{N}_3 e_1 \wedge e_2 \wedge e_6 + \mathbf{N}_4 e_1 \wedge e_2 \wedge e_4.
\end{aligned} \tag{3.65}$$

We can now use these expressions (3.62) and (3.65) to construct the complex 3-form flux $\mathbf{G}_3 = \mathbf{F}_3 - i e^{-\phi} \mathbf{H}_3$. Using the formulas for $\mathbf{F}_3$ and $\mathbf{H}_3$ in (3.2), we have our final expression

for $\mathbf{G}_3$:

$$\begin{aligned}
\mathbf{G}_3 &= e_2 \wedge e_3 \wedge e_4 \left(\mathbf{M}_4 e^{-2\phi}\cosh\gamma + i\mathbf{P}_3 e^{-\phi}\sinh\gamma\right) \\
&+ e_2 \wedge e_5 \wedge e_6 \left(\mathbf{M}_1 e^{-2\phi}\cosh\gamma + i\mathbf{P}_4 e^{-\phi}\sinh\gamma\right) \\
&+ e_2 \wedge e_4 \wedge e_5 \left(\mathbf{M}_2 e^{-2\phi}\cosh\gamma + i\mathbf{P}_1 e^{-\phi}\sinh\gamma\right) \\
&+ e_2 \wedge e_3 \wedge e_6 \left(\mathbf{M}_3 e^{-2\phi}\cosh\gamma + i\mathbf{P}_2 e^{-\phi}\sinh\gamma\right) \\
&+ e_1 \wedge e_4 \wedge e_6 \left(\mathbf{Q}_3 e^{-2\phi}\cosh\gamma + i\mathbf{Q}_2 e^{-\phi}\sinh\gamma\right) \\
&+ e_1 \wedge e_3 \wedge e_5 \left(\mathbf{Q}_4 e^{-2\phi}\cosh\gamma + i\mathbf{Q}_1 e^{-\phi}\sinh\gamma\right) \\
&- e_1 \wedge e_3 \wedge e_6 \left(\mathbf{P}_1 e^{-2\phi}\cosh\gamma - i\mathbf{M}_2 e^{-\phi}\sinh\gamma\right) \\
&- e_1 \wedge e_4 \wedge e_5 \left(\mathbf{P}_2 e^{-2\phi}\cosh\gamma - i\mathbf{M}_3 e^{-\phi}\sinh\gamma\right) \\
&- e_1 \wedge e_5 \wedge e_6 \left(\mathbf{P}_3 e^{-2\phi}\cosh\gamma - i\mathbf{M}_4 e^{-\phi}\sinh\gamma\right) \\
&- e_1 \wedge e_3 \wedge e_4 \left(\mathbf{P}_4 e^{-2\phi}\cosh\gamma - i\mathbf{M}_1 e^{-\phi}\sinh\gamma\right) \\
&- e_2 \wedge e_4 \wedge e_6 \left(\mathbf{Q}_1 e^{-2\phi}\cosh\gamma - i\mathbf{Q}_4 e^{-\phi}\sinh\gamma\right) \\
&- e_2 \wedge e_3 \wedge e_5 \left(\mathbf{Q}_2 e^{-2\phi}\cosh\gamma - i\mathbf{Q}_3 e^{-\phi}\sinh\gamma\right) \\
&+ e_6 \wedge e_3 \wedge e_4 \left(i\mathbf{N}_1 e^{-\phi}\sinh\gamma\right) + e_4 \wedge e_5 \wedge e_6 \left(i\mathbf{N}_2 e^{-\phi}\sinh\gamma\right) \\
&+ e_5 \wedge e_3 \wedge e_4 \left(i\mathbf{N}_3 e^{-\phi}\sinh\gamma\right) + e_3 \wedge e_5 \wedge e_6 \left(i\mathbf{N}_4 e^{-\phi}\sinh\gamma\right) \\
&- e_1 \wedge e_2 \wedge e_5 \left(\mathbf{N}_1 e^{-2\phi}\cosh\gamma\right) - e_1 \wedge e_2 \wedge e_3 \left(\mathbf{N}_2 e^{-2\phi}\cosh\gamma\right) \\
&+ e_1 \wedge e_2 \wedge e_6 \left(\mathbf{N}_3 e^{-2\phi}\cosh\gamma\right) + e_1 \wedge e_2 \wedge e_4 \left(\mathbf{N}_4 e^{-2\phi}\cosh\gamma\right). \quad (3.66)
\end{aligned}$$

We can now use this expression to make the last set of SUSY constraints on our geometry. We do this by requiring $\mathbf{G}_3$ to be a primitive $(2,1)$ form. So we must derive a set of equations that forces all $(3,0), (0,3),$ and $(1,2)$ pieces of $\mathbf{G}_3$ to vanish and then check that the result is primitive. To find the these pieces, we need to write $\mathbf{G}_3$ form in terms of complex vielbeins. The complex vielbeins are a set of linear combinations of the real vielbeins that depend on the complex structure used. We will use the simple complex structure $(i\sigma, i, i)$ defined by

$$E_1 = e_2 + i\sigma e_1, \quad E_2 = e_3 + ie_4, \quad E_3 = e_5 + ie_6 \qquad (3.67)$$

$$\overline{E}_1 = e_2 - i\sigma e_1, \quad \overline{E}_2 = e_3 - ie_4, \quad \overline{E}_3 = e_5 - ie_6, \qquad (3.68)$$

where $\sigma$ is an unspecified real number that we will solve for in 3.4. We first write the $e_i$ basis variables in terms of the $E_i$ basis variables:

$$\begin{aligned}
e_1 &= \frac{E_1 - \overline{E}_1}{2i\sigma}, & e_2 &= \frac{E_1 + \overline{E}_1}{2}, & e_3 &= \frac{E_2 + \overline{E}_2}{2} \\
e_4 &= \frac{E_2 - \overline{E}_2}{2i}, & e_5 &= \frac{E_3 + \overline{E}_3}{2}, & e_6 &\frac{E_3 - \overline{E}_3}{2i}.
\end{aligned} \qquad (3.69)$$

Combining (3.69) and (3.66), yields $\mathbf{G}_3$ in the complex vielbein basis. The result is a large expression whose components we will write out in the next section.

## 3.4 Solving the supersymmetry constraints on $\mathbf{G}_3$

Again, in order for the background to be dual to an $\mathcal{N} = 1$ supersymmetric theory, we require all the $(i, j)$ pieces except for the $(2, 1)$ piece to vanish. An $(i, j)$ piece of $\mathbf{G}_3$, by definition, has $i$ $E_k$ basis vectors wedged together with $j$ $\overline{E}_k$ basis vectors.

First we require the $(3, 0)$ piece of $\mathbf{G}_3$ to vanish. There is only $\binom{3}{3} = 1$ such combination, namely

$$\mathbb{Z}_1 E_1 \wedge E_2 \wedge E_3, \qquad (3.70)$$

where we have named the coefficient in front of the basis 3-form $\mathbb{Z}_1$. We see that in general all of the coefficients $\mathbb{Z}$ in front of each of the terms is complex, so requiring any given $\mathbb{Z}$ to vanish gives two equations, one for its real part Re $\mathbb{Z}$ and one for its imaginary part Im $\mathbb{Z}$.

In particular, for $\mathbb{Z}_1$ we have:

$$\begin{aligned}
\operatorname{Re} \mathbb{Z}_1 &= \frac{\mathbf{P}_1 + \mathbf{P}_2}{8}\left(e^{-\phi}\sinh\gamma + \frac{e^{-2\phi}\cosh\gamma}{\sigma}\right) + \frac{\mathbf{Q}_1 - \mathbf{Q}_2}{8}\left(\frac{e^{-\phi}\sinh\gamma}{\sigma} + e^{-2\phi}\cosh\gamma\right) \\
\operatorname{Im} \mathbb{Z}_1 &= \frac{\mathbf{Q}_3 - \mathbf{Q}_4}{8}\left(e^{-\phi}\sinh\gamma + \frac{e^{-2\phi}\cosh\gamma}{\sigma}\right) - \frac{\mathbf{M}_2 + \mathbf{M}_3}{8}\left(\frac{e^{-\phi}\sinh\gamma}{\sigma} + e^{-2\phi}\cosh\gamma\right).
\end{aligned}$$
(3.71)

These set of equations are too involved to try solving right away, and as we will see there will be simpler equations that will force several variables appearing above to zero. So we will come back to these equations later.

The second piece we want to vanish is the $(0,3)$ piece, which there are again only $\binom{3}{3} = 1$ of:

$$\mathbb{Z}_2 \overline{E}_1 \wedge \overline{E}_2 \wedge \overline{E}_3.$$
(3.72)

The two set of equations we get from $\mathcal{G}_3$ are

$$\begin{aligned}
\operatorname{Re} \mathbb{Z}_2 &= \frac{\mathbf{P}_1 + \mathbf{P}_2}{8}\left(\frac{e^{-2\phi}\cosh\gamma}{\sigma} - e^{-\phi}\sinh\gamma\right) - \frac{\mathbf{Q}_1 - \mathbf{Q}_2}{8}\left(\frac{e^{-\phi}\sinh\gamma}{\sigma} - e^{-2\phi}\cosh\gamma\right) \\
\operatorname{Im} \mathbb{Z}_2 &= \frac{\mathbf{Q}_4 - \mathbf{Q}_3}{8}\left(\frac{e^{-2\phi}\cosh\gamma}{\sigma} - e^{-\phi}\sinh\gamma\right) + \frac{\mathbf{M}_2 + \mathbf{M}_3}{8}\left(e^{-2\phi}\cosh\gamma - \frac{e^{-\phi}\sinh\gamma}{\sigma}\right).
\end{aligned}$$
(3.73)

By the same reasoning as above, we will wait to write down all the equations before attempting to solve any.

Since we have set the $(3,0)$ and $(0,3)$ pieces to zero, all we are missing is to set the $(1,2)$ pieces to zero. However there are $\binom{3}{1}\binom{3}{2} = 9$ different possible combinations of basis 3-forms $E_i \wedge \overline{E}_j \wedge \overline{E}_k$.

For the case $i \neq j \neq k$, we will name the coefficients $\mathbb{Z}_{3i}$, where the $i$ is the subscript of the $E_i$ basis vector. There are three such terms:

$$\mathbb{Z}_{31} E_1 \wedge \overline{E}_2 \wedge \overline{E}_3, \quad \mathbb{Z}_{32} E_2 \wedge \overline{E}_1 \wedge \overline{E}_3, \quad \mathbb{Z}_{33} E_3 \wedge \overline{E}_1 \wedge \overline{E}_2. \tag{3.74}$$

These yield the following six equations:

$$\operatorname{Re} \mathbb{Z}_{31} = \frac{\mathbf{Q}_1 - \mathbf{Q}_2}{8}\left(\frac{e^{-\phi}\sinh\gamma}{\sigma} + e^{-2\phi}\cosh\gamma\right) - \frac{\mathbf{P}_1 + \mathbf{P}_2}{8}\left(\frac{e^{-2\phi}\cosh\gamma}{\sigma} + e^{-\phi}\sinh\gamma\right)$$

$$\operatorname{Im} \mathbb{Z}_{31} = \frac{\mathbf{Q}_3 - \mathbf{Q}_4}{8}\left(\frac{e^{-2\phi}\cosh\gamma}{\sigma} + e^{-\phi}\sinh\gamma\right) + \frac{\mathbf{M}_2 + \mathbf{M}_3}{8}\left(e^{-2\phi}\cosh\gamma + \frac{e^{-\phi}\sinh\gamma}{\sigma}\right)$$

$$\tag{3.75}$$

$$\operatorname{Re} \mathbb{Z}_{32} = \frac{\mathbf{Q}_1 + \mathbf{Q}_2}{8}\left(\frac{e^{-\phi}\sinh\gamma}{\sigma} + e^{-2\phi}\cosh\gamma\right) + \frac{\mathbf{P}_2 - \mathbf{P}_1}{8}\left(\frac{e^{-2\phi}\cosh\gamma}{\sigma} + e^{-\phi}\sinh\gamma\right)$$

$$\operatorname{Im} \mathbb{Z}_{32} = \frac{\mathbf{M}_2 - \mathbf{M}_3}{8}\left(e^{-2\phi}\cosh\gamma + \frac{e^{-\phi}\sinh\gamma}{\sigma}\right) - \frac{\mathbf{Q}_3 + \mathbf{Q}_4}{8}\left(\frac{e^{-2\phi}\cosh\gamma}{\sigma} + e^{-\phi}\sinh\gamma\right).$$

$$\tag{3.76}$$

$$\operatorname{Re} \mathbb{Z}_{33} = \frac{\mathbf{P}_2 - \mathbf{P}_1}{8}\left(\frac{e^{-2\phi}\cosh\gamma}{\sigma} + e^{-\phi}\sinh\gamma\right) - \frac{\mathbf{Q}_1 + \mathbf{Q}_2}{8}\left(\frac{e^{-\phi}\sinh\gamma}{\sigma} + e^{-2\phi}\cosh\gamma\right)$$

$$\operatorname{Im} \mathbb{Z}_{33} = \frac{\mathbf{M}_2 - \mathbf{M}_3}{8}\left(e^{-2\phi}\cosh\gamma + \frac{e^{-\phi}\sinh\gamma}{\sigma}\right) + \frac{\mathbf{Q}_3 + \mathbf{Q}_4}{8}\left(\frac{e^{-2\phi}\cosh\gamma}{\sigma} + e^{-\phi}\sinh\gamma\right).$$

$$\tag{3.77}$$

There is one more case to consider, namely that of $i \neq j, k = i$. We will name the coefficients with $i = k = 2$ as

$$\mathbb{Z}_{41} E_2 \wedge \overline{E}_1 \wedge \overline{E}_2, \quad \mathbb{Z}_{42} E_2 \wedge \overline{E}_3 \wedge \overline{E}_2, \tag{3.78}$$

the case $i = k = 3$ as

$$\mathbb{Z}_{51} E_3 \wedge \overline{E}_1 \wedge \overline{E}_3, \quad \mathbb{Z}_{52} E_3 \wedge \overline{E}_2 \wedge \overline{E}_3, \tag{3.79}$$

and the case $i = k = 1$ as

$$\mathbb{Z}_{61} E_1 \wedge \overline{E}_3 \wedge \overline{E}_1, \quad \mathbb{Z}_{62} E_1 \wedge \overline{E}_2 \wedge \overline{E}_1. \tag{3.80}$$

We get the following twelve equations that are required to vanish:

$$\begin{aligned}
\operatorname{Re} \mathbb{Z}_{41} &= \frac{1}{4} \left( \mathbf{P}_3 e^{-\phi} \sinh \gamma - \frac{\mathbf{P}_4 e^{-2\phi} \cosh \gamma}{\sigma} \right) \\
\operatorname{Im} \mathbb{Z}_{41} &= \frac{1}{4} \left( \frac{\mathbf{M}_1 e^{-\phi} \sinh \gamma}{\sigma} - \mathbf{M}_4 e^{-2\phi} \cosh \gamma \right)
\end{aligned} \tag{3.81}$$

$$\operatorname{Re} \mathbb{Z}_{42} = -\frac{1}{4} \mathbf{N}_3 e^{-\phi} \sinh \gamma, \quad \operatorname{Im} \mathbb{Z}_{42} = \frac{1}{4} \mathbf{N}_1 e^{-\phi} \sinh \gamma \tag{3.82}$$

$$\begin{aligned}
\operatorname{Re} \mathbb{Z}_{51} &= \frac{1}{4} \left( \mathbf{P}_4 e^{-\phi} \sinh \gamma - \frac{\mathbf{P}_3 e^{-2\phi} \cosh \gamma}{\sigma} \right) \\
\operatorname{Im} \mathbb{Z}_{51} &= \frac{1}{4} \left( \frac{\mathbf{M}_4 e^{-\phi} \sinh \gamma}{\sigma} - \mathbf{M}_1 e^{-2\phi} \cosh \gamma \right)
\end{aligned} \tag{3.83}$$

$$\operatorname{Re} \mathbb{Z}_{52} = -\frac{1}{4} \mathbf{N}_4 e^{-\phi} \sinh \gamma, \quad \operatorname{Im} \mathbb{Z}_{52} = \frac{1}{4} \mathbf{N}_2 e^{-\phi} \sinh \gamma \tag{3.84}$$

$$\begin{aligned}
\operatorname{Re} \mathbb{Z}_{61} &= \frac{1}{4\sigma} \mathbf{N}_3 e^{-2\phi} \cosh \gamma, \quad \operatorname{Im} \mathbb{Z}_{61} = \frac{1}{4\sigma} \mathbf{N}_1 e^{-2\phi} \cosh \gamma \\
\operatorname{Re} \mathbb{Z}_{62} &= \frac{1}{4\sigma} \mathbf{N}_4 e^{-2\phi} \cosh \gamma, \quad \operatorname{Im} \mathbb{Z}_{62} = \frac{1}{4\sigma} \mathbf{N}_2 e^{-2\phi} \cosh \gamma.
\end{aligned} \tag{3.85}$$

We see that half of these equations imply that the $\mathbf{N}_i$ variables vanish. So any set of solutions we may find must satisfy that condition.

It turns out the ansatz $\mathbf{A} = -\mathbf{B}$ does set $\mathbf{N}_i = 0$ and makes 18 of the 22 equations vanish. To show this we first compute $\mathcal{F}_1, \mathcal{F}_2$ and $\mathbf{K}$ and their $\psi$ and $r$ derivatives in this limit. From

Chapter 3. Calculating the supersymmetry constraints                                                                  41these quantities we can determine all the $\mathbf{H}$ coefficients and in turn many of the $\mathbf{M}_i, \mathbf{N}_i, \mathbf{P_i}$ and $\mathbf{Q}_i$ coefficients. From (3.53), we see that they simplify to

$$\mathbf{K} = -\mathbf{A}\sqrt{\mathcal{G}_3\mathcal{G}_4}\sin\psi, \quad \mathcal{F}_1 = \mathbf{A}\sqrt{\mathcal{G}_3\mathcal{G}_4}\cos\psi, \quad \mathcal{F}_2 = -\mathbf{A}\sqrt{\mathcal{G}_3\mathcal{G}_4}\cos\psi \qquad (3.86)$$

$$\mathbf{K}_\psi = -\mathbf{A}\sqrt{\mathcal{G}_3\mathcal{G}_4}\cos\psi, \quad \mathcal{F}_{1\psi} = -\mathbf{A}\sqrt{\mathcal{G}_3\mathcal{G}_4}\sin\psi, \quad \mathcal{F}_{2\psi} = \mathbf{A}\sqrt{\mathcal{G}_3\mathcal{G}_4}\sin\psi$$

$$\mathbf{K}_r = -\left(\mathbf{A}\sqrt{\mathcal{G}_3\mathcal{G}_4}\right)_r\sin\psi, \quad \mathcal{F}_{1r} = \left(\mathbf{A}\sqrt{\mathcal{G}_3\mathcal{G}_4}\right)_r\cos\psi, \quad \mathcal{F}_{2r} = -\left(\mathbf{A}\sqrt{\mathcal{G}_3\mathcal{G}_4}\right)_r\cos\psi,$$

where we used the trigonometric identity $\cos^2\frac{\psi}{2} - \sin^2\frac{\psi}{2} = \cos\psi$. We see that (3.86) implies that the $\mathbf{K}_n$ defined in (3.55) must vanish:

$$\mathbf{K}_n = 0, \quad n = 3,\ldots,6. \qquad (3.87)$$

The $\mathbf{K}_n$ vanishing then implies the variables

$$\mathbf{H}_{4i}^{(0)} = 0, \quad \mathbf{H}_{4i}^{'(0)} = 0, \quad i = 1,2,3, \qquad (3.88)$$

defined in (3.60) must vanish. Since the $\mathbf{N}_i$ defined in (3.61) only depend on the $\mathbf{H}_{4i}^{(0)}$ and $\mathbf{H}_{4i}^{'(0)}$, we have

$$\mathbf{N}_i = 0, \quad i = 1,\ldots,4, \qquad (3.89)$$

which was required.

Using (3.86), we can compute the rest of the $\mathbf{H}$ coefficients defined in (3.57), (3.59), and (3.60):

$$\mathbf{H}_{3c}^{(0)} \equiv -\frac{\left(\mathbf{A}\sqrt{\mathcal{G}_3\mathcal{G}_4}\right)_r \sin^2\psi}{\sqrt{\mathcal{G}_1\mathcal{G}_3\mathcal{G}_4}\,(\alpha_1\alpha_3 - \beta_1\beta_3)(\alpha_2\alpha_4 - \beta_2\beta_4)}$$

$$\mathbf{H}_{3a}^{(0)} \equiv \frac{\left(\mathbf{A}\sqrt{\mathcal{G}_3\mathcal{G}_4}\right)_r \sin 2\psi}{2(\alpha_1\alpha_3 - \beta_1\beta_3)\,\mathcal{G}_3\sqrt{\mathcal{G}_1}}, \quad \mathbf{H}_{3b}^{(0)} \equiv -\frac{\left(\mathbf{A}\sqrt{\mathcal{G}_3\mathcal{G}_4}\right)_r \sin 2\psi}{2(\alpha_2\alpha_4 - \beta_2\beta_4)\,\mathcal{G}_4\sqrt{\mathcal{G}_1}},$$

$$\mathbf{H}_{3c}^{(1)} \equiv -\frac{\left(\mathbf{A}\sqrt{\mathcal{G}_3\mathcal{G}_4}\right)_r \cos^2\psi}{\sqrt{\mathcal{G}_1\mathcal{G}_3\mathcal{G}_4}\,(\alpha_1\alpha_3 - \beta_1\beta_3)(\alpha_2\alpha_4 - \beta_2\beta_4)}$$

$$\mathbf{H}_{3a}^{(1)} \equiv -\frac{\left(\mathbf{A}\sqrt{\mathcal{G}_3\mathcal{G}_4}\right)_r \sin 2\psi}{2\left(\alpha_1\alpha_3 - \beta_1\beta_3\right)\mathcal{G}_3\sqrt{\mathcal{G}_1}}, \qquad \mathbf{H}_{3b}^{(1)} \equiv \frac{\left(\mathbf{A}\sqrt{\mathcal{G}_3\mathcal{G}_4}\right)_r \sin 2\psi}{2\left(\alpha_2\alpha_4 - \beta_2\beta_4\right)\mathcal{G}_4\sqrt{\mathcal{G}_1}},$$

$$\mathbf{H}_{3c}^{'(0)} \equiv -\frac{\mathbf{A}\sin 2\psi}{2\sqrt{\mathcal{G}_2}\left(\alpha_1\alpha_3 - \beta_1\beta_3\right)\left(\alpha_2\alpha_4 - \beta_2\beta_4\right)}$$

$$\mathbf{H}_{3a}^{'(0)} \equiv \frac{\mathbf{A}\sqrt{\mathcal{G}_4}\cos^2\psi}{\left(\alpha_1\alpha_3 - \beta_1\beta_3\right)\sqrt{\mathcal{G}_2\mathcal{G}_3}}, \qquad \mathbf{H}_{3b}^{'(0)} \equiv -\frac{\mathbf{A}\sqrt{\mathcal{G}_3}\cos^2\psi}{\left(\alpha_2\alpha_4 - \beta_2\beta_4\right)\sqrt{\mathcal{G}_2\mathcal{G}_4}},$$

$$\mathbf{H}_{3c}^{'(1)} \equiv \frac{\mathbf{A}\sin 2\psi}{2\sqrt{\mathcal{G}_2}\left(\alpha_1\alpha_3 - \beta_1\beta_3\right)\left(\alpha_2\alpha_4 - \beta_2\beta_4\right)}$$

$$\mathbf{H}_{3a}^{'(1)} \equiv \frac{\mathbf{A}\sqrt{\mathcal{G}_4}\sin^2\psi}{\left(\alpha_1\alpha_3 - \beta_1\beta_3\right)\sqrt{\mathcal{G}_2\mathcal{G}_3}}, \qquad \mathbf{H}_{3b}^{'(1)} \equiv -\frac{\mathbf{A}\sqrt{\mathcal{G}_3}\sin^2\psi}{\left(\alpha_2\alpha_4 - \beta_2\beta_4\right)\sqrt{\mathcal{G}_2\mathcal{G}_4}}. \qquad (3.90)$$

Also in the $\mathbf{A} = -\mathbf{B}$ limit, $c_o$ and $b_o$ greatly simplify to

$$c_o = b_o = -1. \qquad (3.91)$$

We can now compute many of the coefficients (3.61) in this limit. Since $\mathbf{Q}_1 = \mathbf{H}_{3a}^{(0)} + \mathbf{H}_{3a}^{(1)}$ and $\mathbf{Q}_1 = \mathbf{H}_{3b}^{(0)} + \mathbf{H}_{3b}^{(1)}$, we see from (3.90) that $\mathbf{H}_{3a}^{(0)} = -\mathbf{H}_{3a}^{(1)}$ and $\mathbf{H}_{3b}^{(0)} = -\mathbf{H}_{3b}^{(1)}$, which implies

$$\mathbf{Q}_1 = \mathbf{Q}_2 = 0. \qquad (3.92)$$

Similarly we have $\mathbf{Q}_3 = \mathbf{H}_{3a}^{'(0)} + \mathbf{H}_{3a}^{'(1)}$ and $\mathbf{Q}_4 = \mathbf{H}_{3b}^{'(0)} + \mathbf{H}_{3b}^{'(1)}$, which from (3.90) simplify to

$$\mathbf{Q}_3 = \frac{\mathbf{A}\sqrt{\mathcal{G}_4}}{\left(\alpha_1\alpha_3 - \beta_1\beta_3\right)\sqrt{\mathcal{G}_2\mathcal{G}_3}}, \qquad \mathbf{Q}_4 = -\frac{\mathbf{A}\sqrt{\mathcal{G}_3}}{\left(\alpha_2\alpha_4 - \beta_2\beta_4\right)\sqrt{\mathcal{G}_2\mathcal{G}_4}}. \qquad (3.93)$$

Since $c_o$ and $b_o$ are equal to $-1$, the equations for the $\mathbf{P}_i$ simplify to a constant times the sum of two $\mathbf{H}$ coefficients. For example, $\mathbf{P}_1$ defined in (3.61) simplifies to $(\alpha_1\alpha_4 + \beta_2\beta_3)(\mathbf{H}_{3c}^{'(0)} + \mathbf{H}_{3c}^{'(1)})$. Staring at (3.90), we see that all the sums of $\mathbf{H}$ variables cancel for the $\mathbf{P}_i$ cases and so we have

$$\mathbf{P}_i = 0, \quad i = 1, \ldots, 4. \qquad (3.94)$$

With all of the $\mathbf{N}_i$, $\mathbf{P}_i$, and $\mathbf{Q}_1, \mathbf{Q}_2$ variables vanishing, we have that the Re $\mathbb{Z}$ equations as well as the Im $\mathbb{Z}_{42}$, Im $\mathbb{Z}_{52}$, Im $\mathbb{Z}_{61}$, Im $\mathbb{Z}_{62}$ equations are automatically satisfied. However, just like the $\mathbf{Q}_3, \mathbf{Q}_4$ variables, cancellations do not occur for the $\mathbf{M}_i$ variables, which evaluate to the following in the $\mathbf{A} = -\mathbf{B}$ limit:

$$\begin{aligned}
\mathbf{M}_1 &\equiv \mathbf{H}_1^{(0)} \alpha_3 \alpha_4 - \mathbf{H}_2^{(0)} \beta_1 \beta_2 - \frac{(\alpha_3 \beta_2 + \alpha_4 \beta_1)\left(\mathbf{A}\sqrt{\mathcal{G}_3 \mathcal{G}_4}\right)_r}{\sqrt{\mathcal{G}_1 \mathcal{G}_3 \mathcal{G}_4}\,(\alpha_1 \alpha_3 - \beta_1 \beta_3)(\alpha_2 \alpha_4 - \beta_2 \beta_4)} \\
\mathbf{M}_2 &\equiv \mathbf{H}_1^{(0)} \alpha_3 \beta_4 - \mathbf{H}_2^{(0)} \beta_1 \alpha_2 - \frac{(\alpha_2 \alpha_3 + \beta_1 \beta_4)\left(\mathbf{A}\sqrt{\mathcal{G}_3 \mathcal{G}_4}\right)_r}{\sqrt{\mathcal{G}_1 \mathcal{G}_3 \mathcal{G}_4}\,(\alpha_1 \alpha_3 - \beta_1 \beta_3)(\alpha_2 \alpha_4 - \beta_2 \beta_4)} \\
\mathbf{M}_3 &\equiv \mathbf{H}_1^{(0)} \alpha_4 \beta_3 - \mathbf{H}_2^{(0)} \beta_2 \alpha_1 - \frac{(\alpha_1 \alpha_4 + \beta_2 \beta_3)\left(\mathbf{A}\sqrt{\mathcal{G}_3 \mathcal{G}_4}\right)_r}{\sqrt{\mathcal{G}_1 \mathcal{G}_3 \mathcal{G}_4}\,(\alpha_1 \alpha_3 - \beta_1 \beta_3)(\alpha_2 \alpha_4 - \beta_2 \beta_4)} \\
\mathbf{M}_4 &\equiv -\mathbf{H}_1^{(0)} \beta_3 \beta_4 + \mathbf{H}_2^{(0)} \alpha_1 \alpha_2 + \frac{(\alpha_1 \beta_4 + \alpha_2 \beta_3)\left(\mathbf{A}\sqrt{\mathcal{G}_3 \mathcal{G}_4}\right)_r}{\sqrt{\mathcal{G}_1 \mathcal{G}_3 \mathcal{G}_4}\,(\alpha_1 \alpha_3 - \beta_1 \beta_3)(\alpha_2 \alpha_4 - \beta_2 \beta_4)}.
\end{aligned} \tag{3.95}$$

We can find some additional constraints on these variables by solving the rest of the SUSY equations that are not already satisfied. The Im $\mathbb{Z}_{41}$ and Im $\mathbb{Z}_{51}$ equations in (3.81) and (3.83) respectively imply the following equations:

$$\begin{aligned}
\frac{\mathbf{M}_4 e^{-\phi} \sinh \gamma}{\sigma} - \mathbf{M}_1 e^{-2\phi} \cosh \gamma &= 0 \\
\frac{\mathbf{M}_1 e^{-\phi} \sinh \gamma}{\sigma} - \mathbf{M}_4 e^{-2\phi} \cosh \gamma &= 0.
\end{aligned} \tag{3.96}$$

We can take $\mathbf{M}_1 = \mathbf{M}_4 = 0$ and leave the parameter $\sigma$ of the complex structure unspecified, or solve for a fixed value $\sigma$ and find a more nontrivial relation for the $\mathbf{M}_i$ variables. Since there is no physical reason to leave $\sigma$ unspecified, we will proceed with the latter. The following solves (3.96):

$$\mathbf{M}_1 = \pm \mathbf{M}_4, \quad \sigma = \pm e^\phi \tanh \gamma. \tag{3.97}$$

So we have

$$\mathbf{M}_1 = -\mathbf{M}_4, \quad \sigma = -e^\phi \tanh \gamma. \tag{3.98}$$

*Chapter 3. Calculating the supersymmetry constraints* 44Furthermore, Im $\mathbb{Z}_1$, Im $\mathbb{Z}_2$, and Im $\mathbb{Z}_{31}$ vanish if we set

$$\mathbf{Q}_3 = \mathbf{Q}_4, \quad \mathbf{M}_2 = -\mathbf{M}_3. \tag{3.99}$$

All that remains is Im $\mathbb{Z}_{32}$ and Im $\mathbb{Z}_{33}$ which, using (3.98) and (3.99), simplify to

$$\text{Im } \mathbb{Z}_{33} = -\text{Im } \mathbb{Z}_{32} = \frac{e^{-\phi}\mathbf{Q}_3}{4}\left(\frac{\sinh^2\gamma - e^{-2\phi}\cosh^2\gamma}{\sinh\gamma}\right). \tag{3.100}$$

These become zero if we set

$$\mathbf{Q}_3 = \mathbf{Q}_4 = 0, \tag{3.101}$$

which since $\mathbf{Q}_3 \propto \mathbf{Q}_4 \propto \mathbf{A}$ (3.93) implies that

$$\mathbf{A} = -\mathbf{B} = 0. \tag{3.102}$$

So our analysis leads to the constraint that $\mathbf{A}$ and $\mathbf{B}$ are equal to zero, which also implies that $\mathbf{A} = \mathbf{B}$, which is why our earlier simplified analysis in the $\mathbf{A} = \mathbf{B}$ limit is completely consistent with our more generalized analysis here.

So all equations demanding that $\mathbf{G}_3$ be a $(2,1)$ form can be compactly described by the following equations:

$$\mathbf{A} = -\mathbf{B} = 0, \quad \mathbf{M}_1 = -\mathbf{M}_4, \quad \mathbf{M}_2 = -\mathbf{M}_3, \tag{3.103}$$

where $\mathbf{A}, \mathbf{B}$ are defined in (3.21), and $\mathbf{M}_1$ and $\mathbf{M}_2$ simplify to

$$\begin{aligned}\mathbf{M}_1 &= \mathbf{H}_1^{(0)}\alpha_3\alpha_4 - \mathbf{H}_2^{(0)}\beta_1\beta_2 \\ \mathbf{M}_2 &= \mathbf{H}_1^{(0)}\alpha_3\beta_4 - \mathbf{H}_2^{(0)}\beta_1\alpha_2.\end{aligned} \tag{3.104}$$

The final expression for $\mathbf{G}_3$ then reads

$$\frac{\mathbf{G}_3}{\cosh \gamma} = \frac{ie^{-2\phi}\mathbf{M}_1}{2} E_1 \wedge \left(E_3 \wedge \overline{E}_3 - E_2 \wedge \overline{E}_2\right)$$
$$+ \frac{ie^{-2\phi}\mathbf{M}_2}{2} E_1 \wedge \left(\overline{E}_2 \wedge E_3 - E_2 \wedge \overline{E}_3\right). \quad (3.105)$$

All that is left to check is that $\mathbf{G}_3$ is primitive, that is it satisfies $J \wedge \mathbf{G}_3 = 0$. Using

$$J = -\frac{i}{2}\left(E_1 \wedge \overline{E}_1 + E_2 \wedge \overline{E}_2 + E_3 \wedge \overline{E}_3\right), \quad (3.106)$$

we have

$$J \wedge \mathbf{G}_3 = \frac{e^{-2\phi}\mathbf{M}_1}{4}\left(E_2 \wedge \overline{E}_2 \wedge E_1 \wedge E_3 \wedge \overline{E}_3 - E_3 \wedge \overline{E}_3 \wedge E_1 \wedge E_2 \wedge \overline{E}_2\right) = 0 \quad (3.107)$$

by the commutative properties of the wedge product. So we have that $\mathbf{G}_3$ is primitive.

# Chapter 4

# Conclusion

Beginning with the general metric ansatz describing the non-Kähler resolved warped-deformed conifold in (3.4), we managed to find constraints on the warp factors $\mathcal{G}_i$ so that the supergravity background (3.2) is dual to an $\mathcal{N} = 1$ SUSY gauge theory. We did this by demanding that the metric describes a complex manifold and demanding the complex 3-form $\mathbf{G}_3$ be a primitive $(2, 1)$ form. These generalize the constraints found in [4] where the non-Kähler warped-resolved conifold was studied instead.

The generality of this background hints at a continuous set of new gauge-gravity dualities, however more analysis needs to be done to connect these supergravity theories to their respective dual gauge theories. At the very least, the constraints derived can be used as a test for $\mathcal{N} = 1$ SUSY for brane/flux setups whose supergravity limit lives in the space of solutions defined in (3.4). Further research along this direction could involve relaxing the assumptions (3.5), that is, considering the even more general cases satisfying

$$\mathcal{G}_i = \mathcal{G}_i(r, \theta_1, \theta_2, \phi_1, \phi_2, \psi), \quad i = 1, \ldots, 6,$$

$$\mathcal{G}_5 \neq \mathcal{G}_6. \tag{4.1}$$